\documentclass[aps,amsmath,amssymb,prd,showpacs,floatfix,preprint,superscriptaddress,nofootinbib,12pt]{article}
\usepackage{jheppub}
\usepackage{slashed,verbatim}
\pdfoutput=1
\usepackage{fancyhdr}
\usepackage{epigraph}
\makeatletter
\def\@fpheader{\relax}
\makeatother

\usepackage{lipsum}

\newcommand\blfootnote[1]{%
  \begingroup
  \renewcommand\thefootnote{}\footnote{#1}%
  \addtocounter{footnote}{-1}%
  \endgroup
}

\usepackage{gensymb}
\usepackage{bbding}
\usepackage{pifont}
\usepackage{wasysym}
\usepackage{amssymb}

\usepackage[svgnames]{xcolor}
\usepackage[framemethod=tikz]{mdframed}

\usepackage{amsmath,epsfig}
\usepackage{amssymb,amsfonts}
\usepackage{latexsym}
\usepackage{graphicx}
\usepackage{subeqnarray}
\usepackage{xcolor}

\usepackage{graphicx}
\usepackage{ulem}

\newcommand{\cool}{\ensuremath{%
  \mathchoice{\includegraphics[height=2ex]{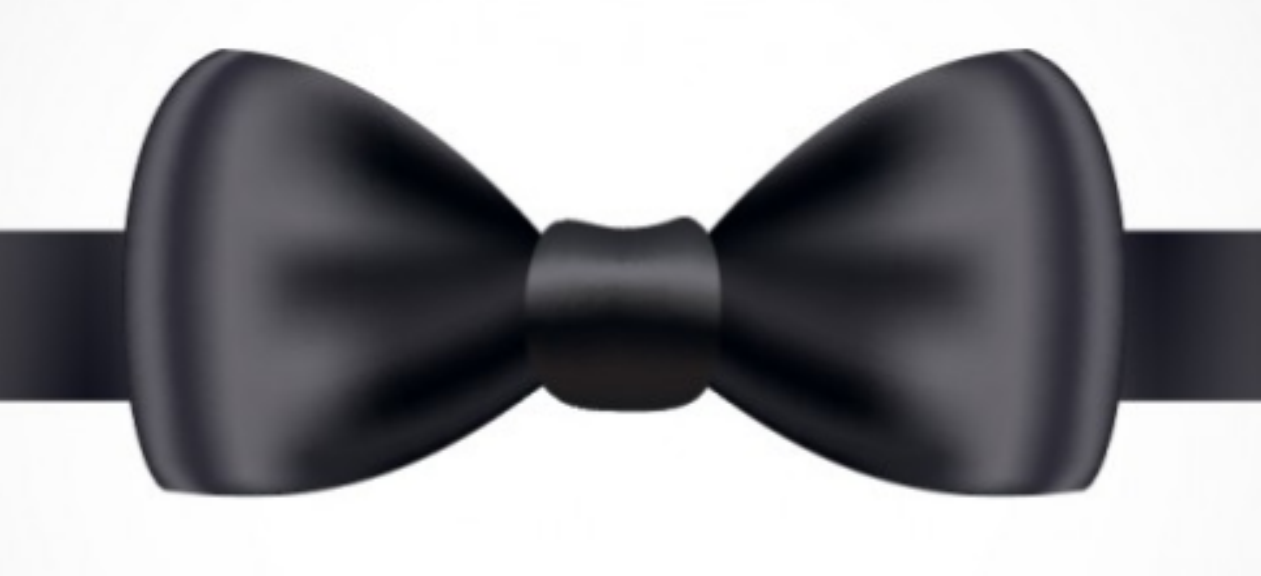}}
    {\includegraphics[height=2ex]{png2pdf.pdf}}
    {\includegraphics[height=1.5ex]{png2pdf.pdf}}
    {\includegraphics[height=1ex]{png2pdf.pdf}}
}}

\newcommand{\me}{\ensuremath{%
  \mathchoice{\includegraphics[height=2ex]{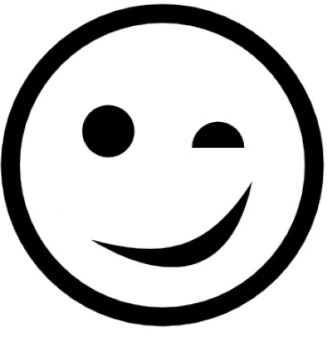}}
    {\includegraphics[height=2ex]{cat.png}}
    {\includegraphics[height=1.5ex]{cat.png}}
    {\includegraphics[height=1ex]{cat.png}}
}}

\def\be{\begin{equation}}
\def\ee{\end{equation}}
\def\bea{\begin{eqnarray}}
\def\eea{\end{eqnarray}}

\newcommand\fverb{\setbox\pippobox=\hbox\bgroup\verb}
\newcommand\fverbdo{\egroup\medskip\noindent%
                        \fbox{\unhbox\pippobox}\ }
\newcommand\fverbit{\egroup\item[\fbox{\unhbox\pippobox}]}

\newcommand{\bear}{\begin{eqnarray}}

\newcommand{\eear}{\end{eqnarray}}

\newcommand{\bsea}{\begin{subeqnarray}}
\newcommand{\esea}{\end{subeqnarray}}
\newbox\pippobox

\def\6{\partial}

\newcommand{\comments}[1]{}

\input Starburst.fd

\allowdisplaybreaks[3]

\preprint{IFT-UAM/CSIC-19-48}

\begin{document}

\title{\centering \huge A Unified Description of Translational Symmetry Breaking in Holography}

\author[\,\,\Diamond]{Martin Ammon}
\affiliation[\,\,\Diamond]{Theoretisch-Physikalisches Institut, Friedrich-Schiller-Universit¨at Jena, Max-Wien-Platz 1, D-07743 Jena, Germany.}

\author[\,\,\cool, \,\me]{, Matteo Baggioli}
\affiliation[\cool]{Instituto de Fisica Teorica UAM/CSIC,
c/ Nicolas Cabrera 13-15, Cantoblanco, 28049 Madrid, Spain}

\author[\,\,\cool]{, Amadeo Jimenez Alba}

\vspace{1cm}

\emailAdd{martin.ammon@uni-jena.de}
\emailAdd{matteo.baggioli@uam.es}
\emailAdd{amadeo.j@gmail.com}

\blfootnote{\me   \,\,\,\url{https://members.ift.uam-csic.es/matteo.baggioli}}

\vspace{1cm}

\abstract{We provide a complete and unified description of translational symmetry breaking in a simple holographic model. In particular, we focus on the distinction and the interplay between explicit and spontaneous breaking. We consider a class of holographic massive gravity models which allow to range continuously from one situation to the other. We study the collective degrees of freedom, the electric AC conductivity and the shear correlator in function of the explicit and spontaneous scales. We show the possibility of having a sound-to-diffusion crossover for the transverse phonons. Within our model, we verify the validity of the Gell-Mann-Oakes-Renner relation. Despite of strong evidence for the absence of any standard dislocation induced phase relaxation mechanism, we identify a novel relaxation scale controlled by the ratio between the explicit and spontaneous breaking scales. Finally, in the pseudo-spontaneous limit, we prove analytically the relation, which has been discussed in the literature, between this novel relaxation scale, the mass of the pseudo-phonons and the Goldstone diffusivity. Our numerical data confirms this analytic result.}

\maketitle

\section{Introduction}
\epigraph{There are no perfect symmetries, there is no pure randomness, we are in the gray region between truth and chaos.
Nothing novel or interesting happens unless it is on the border between order and chaos.}{\textit{R.A.Delmonico}}
Symmetries are elegant principles which constitute the fundamental pillars of many physical theories \cite{stewart2007beauty}. Symmetries are widely accepted beauty standards and they are present in very diverse natural creations ranging from cauliflowers to seashells \cite{Enquist1994}. However their breaking at low energy is ubiquitous. Motivated by important problems in condensed matter and  particle physics, a lot of progress and effort have been devoted to the description and understanding of the breaking patterns of internal global symmetries \cite{coleman1988aspects}. Paradigmatic examples are certainly the cases of the Pion \cite{PhysRevLett.29.1698} or the Higgs mechanism \cite{bhattacharyya2011pedagogical}.

The situation regarding spacetime symmetries appears to be more complicated and is still not well understood \cite{low2002spontaneously,kharuk2018goldstone,kharuk2018solving,HAYATA2014195}. Nevertheless, almost all the condensed matter systems we know in nature break both rotational and translational invariance. Understanding the breakdown patterns of spacetime symmetries represents a valuable problem beyond the academic interest. As a matter of fact, these have a deep influence on the thermodynamic and transport properties of the materials, e.g. the phonons in the context of specific heat.

Much can be learned from the symmetry breaking pattern for a specific system. In particular, powerful effective techniques allow us to classify the various phases of matter accordingly indeed to their broken symmetries \cite{Nicolis:2015sra}. From a more mathematical perspective, this accounts for a robust coset construction of these various possibilities \cite{Nicolis:2013lma}. Moreover, the effective hydrodynamic description can be endowed to account for non-hydro damped modes and/or the possible spontaneous breaking of certain global symmetries. Specific examples of such generalization are the framework of generalized hydrodynamics \cite{boon1991molecular}, quasi-hydrodynamics \cite{Grozdanov:2018fic,Baggioli:2019jcm} or just superfluid hydrodynamics \cite{Davison:2016hno}. Certainly, the same extension can be applied to spacetime symmetries like translations, which are essential to distinguish solids from fluids \cite{PhysRevB.22.2514,PhysRevA.6.2401,Delacretaz:2017zxd}.

\begin{figure}
    \centering
    \includegraphics[width=8.5cm]{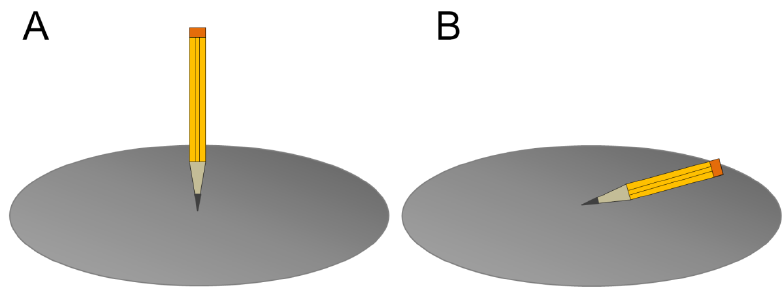}
    \quad 
    \includegraphics[width=6.5cm]{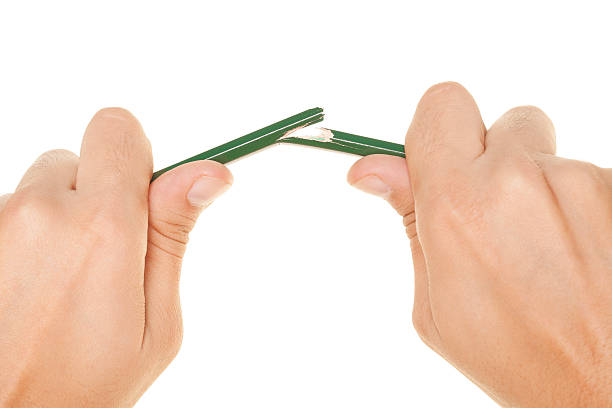}
    \caption{A cartoon of the difference between explicit and spontaneous breaking of a symmetry. \textbf{Left: }The symmetry is broken dynamically by the evolution of the system, which chooses a ground state not respecting such fundamental invariance. \textbf{Right: }The symmetry is broken explicitly by an external source.}
    \label{sk}
\end{figure}

One fundamental point for the following discussions concerns the difference between the explicit breaking (EXB) and the spontaneous breaking (SSB) of a symmetry (see fig. \ref{sk} for a cartoon). From a theoretical point of view, this distinction is very neat. By explicit breaking we mean the breakdown of the symmetry at the fundamental microscopic level, which appears directly in the action of the theory. This is usually caused by an external source that controls the strength of the explicit breaking. One simple example is the application of an external magnetic field to a material. In contrast, the spontaneous symmetry breaking is a dynamical mechanism that is only manifest in the IR physical regime of the theory. In particular the dynamical choice of a ground state which does not respect the fundamental symmetries of the system. A crucial difference is that SSB is accompanied by the appearance of Goldstone modes, \textit{i.e.} new emergent massless degrees of freedom which parametrize the space of equivalent ground states. Following the example of the magnetic field, its spontaneous counterpart is the spontaneous magnetization in a ferromagnetic material. 

This picture is more complicated in the case of translational invariance. The explicit breaking of translations can be, for example, identified as an immobile and non dynamical ionic lattice through which the electrons are scattered dissipating their momentum. At the same time, the dynamical formation of a lattice structure, as it happens for charge density waves \cite{RevModPhys.60.1129}, is described by spontaneous breaking of translations. In simple words the explicit breaking of translations relates directly to the non-conservation of the momentum for the electrons while the SSB accounts for the elastic properties of the system and the propagation of sound.

In this manuscript we will make use of holographic and hydrodynamic methods to better understand the spontaneous and pseudo-spontaneous breaking of translational invariance along with its consequences on the spectrum of excitations and on electric transport. We will also provide a unified picture of the breakdown of translations in a simple holographic model. This work builds on our previous results contained in \cite{Baggioli:2014roa,Alberte:2017cch,Alberte:2017oqx}.\\[0.5cm]
\textbf{Spontaneous symmetry breaking of translations and phonons}\\[0.1cm]
The SSB of translational invariance is a common situation in real materials, which appears whenever an ordered structure is dynamically formed. Typical examples are the lattice of ordered crystals \cite{kosevich2006crystal} and the formation of charge density waves or striped phases \cite{RevModPhys.60.1129}. In both cases, long-range order is established in the sample and it is accompanied by the presence of massless Goldstone bosons, which in the case of the Ionic lattice are known as phonons\footnote{In the context of CDW the Goldstone excitations are referred to as \textit{sliding modes} or \textit{phasons}. Their nature is similar to that of acoustic phonons.}. More specifically, acoustic phonons\footnote{Optical phonons have a finite and usually large mass gap and they cannot be interpreted as Goldstone modes for translations.} can be indeed thought of as the Goldstone bosons for translational invariance \cite{leutwyler1996phonons}. Their dispersion relation at low energy is, as expected for Goldstone bosons, of the form:
\begin{equation}
    \omega_T\,=\,c_T\,k\,+\,\dots \, ,
\,,\quad \omega_L\,=\,c_L\,k\,+\,\dots\, ,
\end{equation}
where $T,L$ stand for transverse and longitudinal with respect to momentum $\vec{k}$. The speeds (in $2+1$ dimensions) are related to the elastic shear and bulk moduli $G,K$ of the material accordingly by
\begin{equation}
    c_T^2\,=\,\frac{G}{\chi_{PP}}\,,\quad c_L^2\,=\,\frac{G\,+\,K}{\chi_{PP}}\, ,
\end{equation}
as derived in standard textbooks \cite{landau7,Lubensky}. Here we denote by $\chi_{PP}$ the momentum static susceptibility.

The situation is more complicated when dissipative effects have to be taken into account. This is indeed the case for highly disordered systems \cite{allen1999diffusons}, viscoelastic materials where the elastic properties coexist with viscous contributions and even ordered crystals with anharmonic effects \cite{landau2}. In those scenarios the dispersion relations of the phonons read
\begin{equation}
    \omega^2\,=\,\tilde{\Omega}_{T,L}^2(k)\,-\,i\,\omega\,\tilde{\Gamma}_{T,L}(k) \, ,
\end{equation}
where $\Gamma_{T,L}$ is the damping coefficient, which incorporates the effects of dissipation. At low momenta we expand $\tilde{\Omega}_{T,L}$ and $\tilde{\Gamma}_{T,L}$ into a power series in $k$,
\begin{equation}
    \tilde{\Omega}_{T,L}(k)\,=\,c_{T,L}\,k\,+\,\dots\,,\quad \tilde{\Gamma}_{T,L}(k)\,=\,D_{T,L}\,k^2\,+\,\dots \, ,
\end{equation}
where $D_{T,L}$ are the diffusion constants, sometimes also called sound attenuation constants. The fact that the leading term of the damping is quadratic in the momentum, \textit{i.e.} diffusive, it is found both experimentally and by microscopic computations \cite{Lubensky}. 

The immediate effects of the dissipative terms, independently of their microscopic origin, is that the phonons are no longer propagating untill arbitrary large values of energy/momenta. At this scale, the dissipative terms become dominant and the phonon modes become diffusive. In certain contexts, after this crossover, the diffusive modes are called \textit{diffusons} \cite{allen1999diffusons}. The crossover between ballistic sound propagation and the diffusive incoherent regime is usually referred to as the Ioffe-Regel crossover \cite{PhysRevB.61.12031,taraskin2002vector,beltukov2013ioffe} and it appears when the mean free path of the phonons is of order of their wavelength. The diffusive modes discussed are quasi-localized and they become totally localized (their diffusion constant goes to zero) at the mobility-edge beyond which the modes are defined \textit{locons} \cite{allen1999diffusons}.\\

For the rest of the paper we focus on the transverse sector of the system.
Massless phonons with linear dispersion relation, obeying the expressions mentioned above, have been identified in holographic massive gravity models in \cite{Alberte:2017oqx}. Because of space limitations, the analysis therein has been limited to the low momentum hydrodynamic limit $k/T\ll 1$ and the Ioffe-Regel crossover has not been analyzed. In this manuscript we continue the analysis beyond the hydrodynamic limit and we show that at high enough temperature the phonons indeed stop propagating and become purely diffusive. Moreover, whenever the crossover happens within the hydrodynamic limit $k/T \ll 1$, we check explicitly that the hydrodynamic expectations \cite{Amoretti:2018tzw}
\begin{equation}
    \omega_{\pm}\,=\,-\,\frac{i}{2}\,k^2\,\left(\xi\,+\,\frac{\eta}{\chi_{PP}}\right)\,\pm\,k\,\sqrt{\frac{G}{\chi_{PP}}\,-\,\frac{k^2}{4}\,\left(\frac{\eta}{\chi_{PP}}\,-\,\xi \right)^2}\,+\,\dots \, ,
\end{equation}
are satisfied. Here, $G$ is the shear modulus, $\eta$ the shear viscosity and $\xi$ the Goldstone diffusion constant.\footnote{As we will explain in the main text of the paper, the terms of order $k^4$ (or higher) have to be taken with a grain of salt.}

Moreover, we also verify that the crossover appears at larger momenta when the temperature is increased. Finally, at a certain critical temperature the crossover does not appear anymore and the low energy phonons connect directly to the high energy UV excitations $\omega=k$. See section \ref{sec: spont} for more details.\\[0.2cm]
\textbf{Pseudo-spontaneous breaking of translations and pseudo-phonons}\\[0.1cm]
Of particular interest is also the situation where the SSB is accompanied by a small explicit breaking source. This is the case in the presence of disorder or impurities, and hence it is relevant in many condensed matter situations. In \cite{Delacretaz:2016ivq} the interplay between SSB and explicit breaking (EXB) has been advocated as a possible explanation for the bad metals transport properties. As a consequence, it is important to investigate such interplay within scenarios at strong coupling lacking a quasiparticle description, for example within holography. The coexistence of SSB and EXB has already been introduced in several holographic models \cite{Jokela:2017ltu,Andrade:2017cnc,Alberte:2017cch,Amoretti:2018tzw}. In this manuscript we analyze the nature of this phenomenon in detail both from the point of view of the spectra of modes and from the optical transport features.

The first observation, originally presented in \cite{Alberte:2017cch}, is that the existence of weakly gapped and damped phonons is the consequence of the collision of two imaginary modes which appear as solutions of the following expression
\begin{equation}
    \left(\bar{\Omega}\,-\,i\,\omega\right)\,\left(\Gamma\,-\,i\,\omega\right)\,+\,\omega_0^2\,=\,0 \, ,
\end{equation}
where $\Gamma$ is the momentum relaxation rate and $\omega_0$ the pinning frequency. The interesting new aspect is what we define with the symbol $\bar{\Omega}$. This corresponds to a non-hydrodynamic mode\footnote{See also \cite{Baggioli:2018vfc,Baggioli:2018nnp,Grozdanov:2018fic} for similar situations where the non-hydrodynamic modes play a fundamental role.} that in the limit of pseudo-spontaneous breaking becomes underdamped and enters in the hydrodynamic window $\omega/T,k/T \ll 1$. It collides with the Drude pole $\omega=-i \Gamma$, which appears underdamped because of the small EXB, and it produces together with the latter the gapped phonon. As we will show in the following, this mechanism is completely analogous to the coherent-incoherent transition appearing in the case of purely explicit breaking \cite{Davison:2014lua,Kim:2014bza}. The only, but crucial, difference is that this time the collision happens at very low frequencies, within the hydrodynamic limit $\omega/T \ll 1$.

In \cite{Amoretti:2018tzw} the nature of the parameter $\bar{\Omega}$ has been attributed to phase relaxation and fluctuating order. As explained in \cite{Delacretaz:2017zxd} a possible microscopic mechanism giving rise to this effect is the proliferation of dislocations and topological defects. As we will show in our paper, the parameter $\bar{\Omega}$ appearing in our holographic model cannot be linked with phase relaxation \textit{stricto sensu}. Its nature is not given by the dynamics of topological defects nor by the fluctuations of the spontaneous order parameter. The reasons why can be summarized as follows:
\begin{itemize}
    \item The frequency dependent viscosity obtained from the shear correlator in the purely SSB limit does not present a Drude peak structure but a simple pole $\sim i/\omega$. Following the hydrodynamic results of \cite{Delacretaz:2017zxd} this implies $\Omega=0$ in our model.
    \item The longitudinal Goldstone diffusive mode is not gapped at zero explicit breaking but it follows a simple dispersion relation $\omega=-i D_\phi k^2$.
    \item The parameter $\bar{\Omega}$ depends on the explicit breaking scale and it vanishes with it.
\end{itemize}
Moreover, we do not find evidence for a phase relaxation mechanism provided by dislocations or topological configurations in our holographic model. This is in agreement with no large diffusivities.
That said, it is very interesting to understand what is the nature of the $\bar{\Omega}$ parameter seen in our model and in \cite{Amoretti:2018tzw,Andrade:2018gqk}. What appears to be clear from all these results is that this parameter depends on the EXB and it is not captured\footnote{This was already suggested in \cite{Andrade:2018gqk}.} by the hydrodynamic description of \cite{Delacretaz:2017zxd}. We hope that our results help to complete this picture\footnote{The presence of a ``phase relaxation'' mechanism given by disorder is known in the condensed matter community \cite{doi:10.1143/JPSJ.45.1474,fogler2000dynamical} and it can probably explain the $\bar{\Omega}$ parameter observed in holography. We thank Blaise Gouteraux to suggest us this point.}.

In the rest of the manuscript we will try to shed some light on the nature of such parameter in a simple holographic massive gravity model. At the same time, we will study in detail the hydrodynamic behaviour of the system and the crossover between explicit and spontaneous breaking. We will moreover determine the dependence of the various parameters in terms of the explicit and spontaneous breaking scales and prove the expected Gell-Mann-Oakes-Renner (GMOR) relation \cite{PhysRev.175.2195}.
Finally we will prove analitically the relation proposed in \cite{Amoretti:2018tzw} between the relaxation scale $\bar{\Omega}$, the mass of the pseudo-Goldstone modes and the Goldstone diffusivity.
\section{The holographic model}\label{sec: model}
We consider the simple holographic massive gravity models introduced in  \cite{Baggioli:2014roa,Alberte:2015isw} and defined by the following action: 
\begin{equation}\label{action}
S\,=\, M_P^2\int d^4x \sqrt{-g}
\left[\frac{R}2+\frac{3}{\ell^2}- \, m^2 V(X)\,-\,\frac{1}{4}\,F^2\right]\, ,
\end{equation}
with $X \equiv \frac12 \, g^{\mu\nu} \,\partial_\mu \phi^I \partial_\nu \phi^I$ and $F^2=F_{\mu\nu}F^{\mu\nu}$ with the field strength being $F=dA$. The dual field theory was studied in detail in \cite{Baggioli:2015zoa,Baggioli:2015gsa,Baggioli:2015dwa,Alberte:2016xja,Alberte:2017cch,Alberte:2017oqx,Andrade:2019zey}. These models are particularly suitable to implement the breaking of translational invariance within the holographic bottom-up framework\footnote{See \cite{Grozdanov:2018ewh,Amoretti:2017frz} for other alternatives.}. We study 4D asymptotically AdS black hole geometries in Eddington-Finkelstein (EF) coordinates:
\begin{equation}
\label{backg}
ds^2=\frac{1}{u^2} \left[-f(u)\,dt^2-2\,dt\,du + dx^2+dy^2\right]\, ,
\end{equation}
where $u\in [0,u_h]$ is the radial holographic direction spanning from the boundary to the horizon, defined through $f(u_h)=0$. The $\phi^I$ are the St\"uckelberg scalars which admit  a radially constant profile $\phi^I=x^I$ with $I=x,y$ and $A_t= \mu\,-\,\rho\, u$ is the background solution for the $U(1)$ gauge field encoding the chemical potential $\mu$ and the charge density $\rho=\mu/u_h$ (for more details see \cite{Alberte:2015isw,Alberte:2017cch}). The emblackening factor takes the simple form:
\begin{equation}\label{backf}
f(u)= u^3 \int_u^{u_h} dv\;\left[ \frac{3}{v^4} -\frac{m^2}{v^4}\, 
V(v^2)\,-\,\frac{\rho^2}{2} \right] \, .
\end{equation}
The corresponding temperature of the dual QFT reads:
\begin{equation}
T=-\frac{f'(u_h)}{4\pi}=\frac{6 -  2 m^2 V\left(u_h^2 \right)\,-\,\mu^2\,u_h^2 }{8 \pi u_h}\, ,\label{eq:temperature}
\end{equation}
while the entropy density is simply $s=2\pi/u_h^2$. The heat capacity can be simply obtained as $c_v=T ds/dT$ and was studied in \cite{Baggioli:2015gsa,Baggioli:2018vfc}.\\

\begin{figure}[hbt]
    \centering
    \includegraphics[width=11cm]{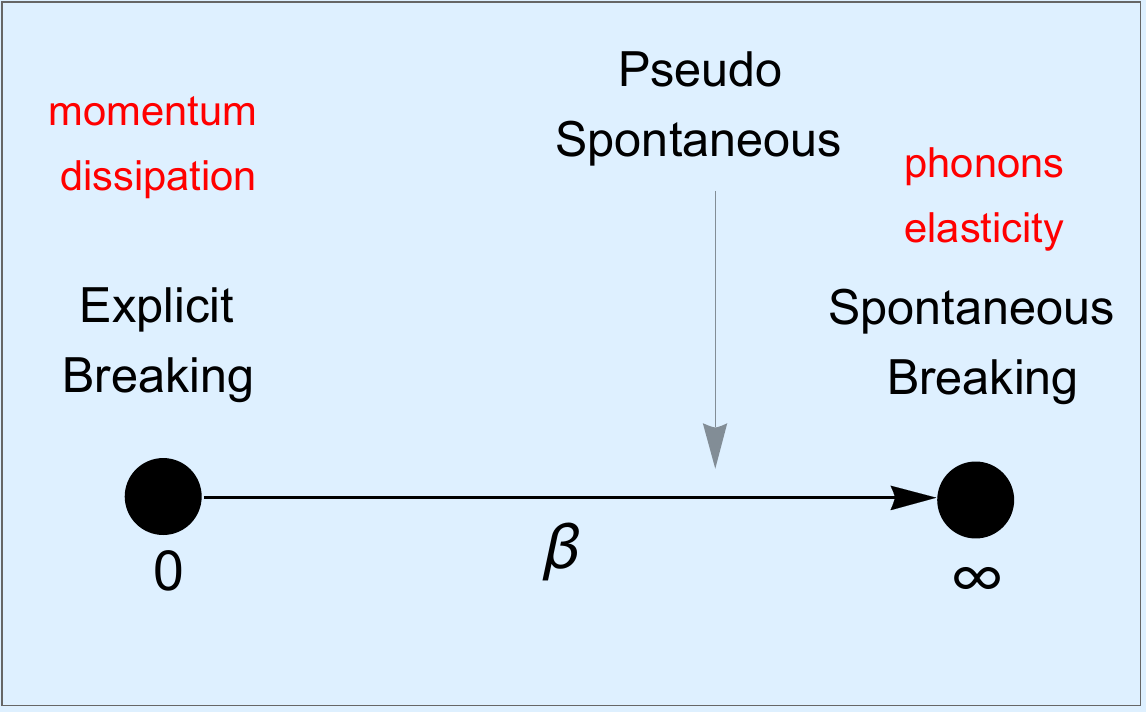}
    \caption{A cartoon for the benchmark model \eqref{benchmark}. The parameter $\beta$ allow to continuously move from an explicit breaking situation towards a purely spontaneous one. Most of the computations of this work will focus on the pseudo-spontaneous regime defined as the limit $\beta \gg 1$.}
    \label{figmod}
\end{figure}

Given the previously mentioned solution for the background scalars, this model allows for explicit or spontaneous breaking of translational invariance depending of the choice of the potential $V(X)$ \cite{Alberte:2017oqx}. In particular if we consider a simple monomial form $V(X)=X^N$ we have:
\begin{equation}
\begin{cases}
1/2\,\leq\,N\,<\,5/2\,\quad \quad \text{EXPLICIT BREAKING (EXB)}\\
N\,>\,5/2\,\quad \quad \hspace{1.35cm}\text{SPONTANEOUS BREAKING (SSB)}\\
\end{cases}
\end{equation}
The simplest example of explicit breaking can be found in the original paper \cite{Andrade:2013gsa} while the spontaneous case is analyzed in detail in \cite{Alberte:2017oqx}. Moreover one can combine the two cases and consider a potential of the form:
\begin{equation}
    V(X)\,=\,X\,+\,\beta\,X^N\,,\quad N\,>5/2\, ,\label{benchmark}
\end{equation}
This is going to be our choice for most of the manuscript.
As shown in fig. \ref{figmod} this choice is convenient because it interpolates continuously between the explicit breaking case to the spontaneous one. For the rest of the manuscript we assume the explicit breaking scale to be small compared to the temperature of the system. In that limit, it was shown in \cite{Alberte:2017cch} that whenever the parameter $\beta$ is large enough the aforementioned potential realize the pseudo-spontaneous breaking of translations introducing a small gap to the would be massless Goldstone bosons\footnote{See \cite{Andrade:2017cnc,Jokela:2017ltu,Amoretti:2018tzw,Donos:2019tmo,Li:2018vrz,Filios:2018xvy,Amoretti:2016bxs,Musso:2018wbv,Donos:2018kkm} for other studies of the interplay of spontaneous and explicit breaking of translations in holography and field theory.}. Moreover in the limit of $\beta \rightarrow \infty$ we can neglect the first term in the potential and we reproduce the results of \cite{Alberte:2017oqx}.\\[0.3cm]
\textbf{Some technicalities}\\
Before proceeding to the next section let us make some technical remarks about this model. The model has been introduced and discussed in details in \cite{Baggioli:2014roa,Alberte:2015isw,Alberte:2017oqx}; we will just repeat some of the points here to help the reader.
\begin{itemize}
    \item[(I)] The reason why the source for the scalar operator $m$ seems to appear in the bulk action in the parameter is due to the normalization choice. A more orthodox way of introducing a source for the scalar following the holographic dictionary would be to consider a solution $\phi^I=s\,x^I$ where in this case $s x^I=\phi_0(x^I)$ is clearly a boundary source. Notice that the parameters $m$ and $s$ are not independent. In this way we can get rid off the parameter $m$ in the action easily. Nevertheless, for simplicity we prefer to keep these notations.
    \item[(II)] In presence of an explicit source for the dual operator (which is the case for a potential $V(X)=X+\beta X^N$), there is no question about the thermodynamic stability of the solution. In the spontaneous case considered in \cite{Alberte:2017cch} the question is more subtle. The solution which naively minimizes the Free energy is the one with zero VEV $\phi^I=0$. Nevertheless the theory we are using cannot be defined around such a solution because of the problems of strong coupling and it makes sense only around a non trivial vacuum. The consistency of our solution is confirmed by the absence of any unstable quasinormal mode (with positive imaginary part).
\end{itemize}

\section{A spontaneous digression}\label{sec: spont}
Before proceeding to analyze the case of the pseudo-spontaneous breaking of translational invariance, let us consider the spontaneous symmetry breaking (SSB) of translations which was originally analyzed in \cite{Alberte:2017oqx}. In order to do that, we set for simplicity the chemical potential to zero, $\mu=0$, and we restrict ourselves to the potentials of the form\footnote{This corresponds to take the benchmark potential \eqref{benchmark} we will use in the rest of the paper and set $m^2=0$ while keeping $m^2 \beta$ finite.}:
\begin{equation}
    V(X)\,=\,X^N\,,\quad N\,>\,\frac{5}{2}\label{ch}\, ,
\end{equation}
for which transverse massless phonons are present in the QNMs spectrum\footnote{It is important to remark that, choosing the potential \eqref{ch}, the background solution with $X=0$ (which corresponds to unbroken translations) is not reliable because of strong coupling. In other words, one cannot immediately compare the solution with unbroken translations with the one where SSB happens, \textit{i.e.} $X\neq 0$. This point has been explained in \cite{Alberte:2017oqx}.} \cite{Alberte:2017oqx}. Notice the analogies with the effective field theories for the spontaneous breaking of the Poincar\'e group and elasticity \cite{Nicolis:2015sra,Alberte:2018doe}. 

\begin{figure}
    \centering
    \includegraphics[width=7.5cm]{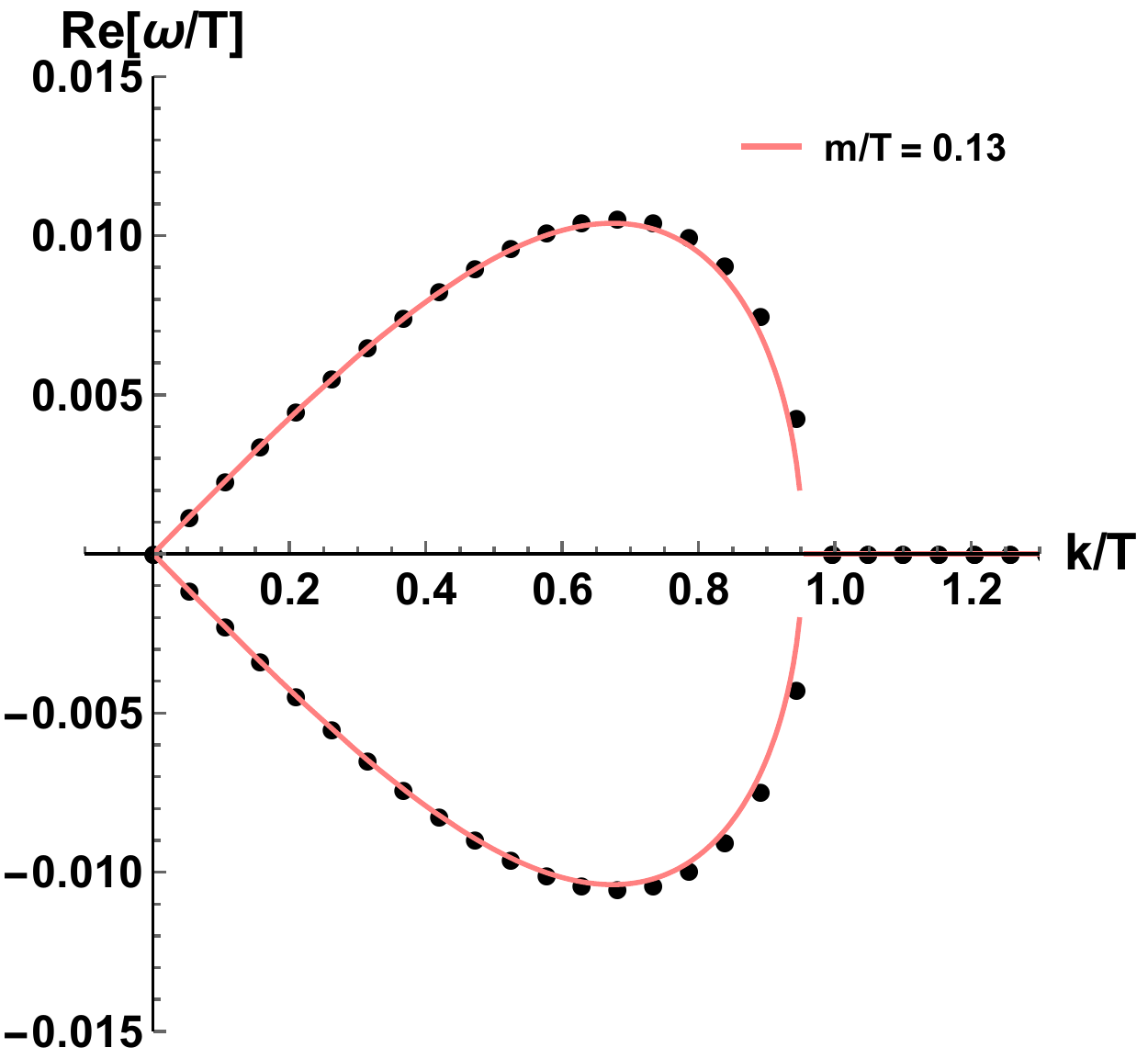}
    \quad
    \includegraphics[width=7.5cm]{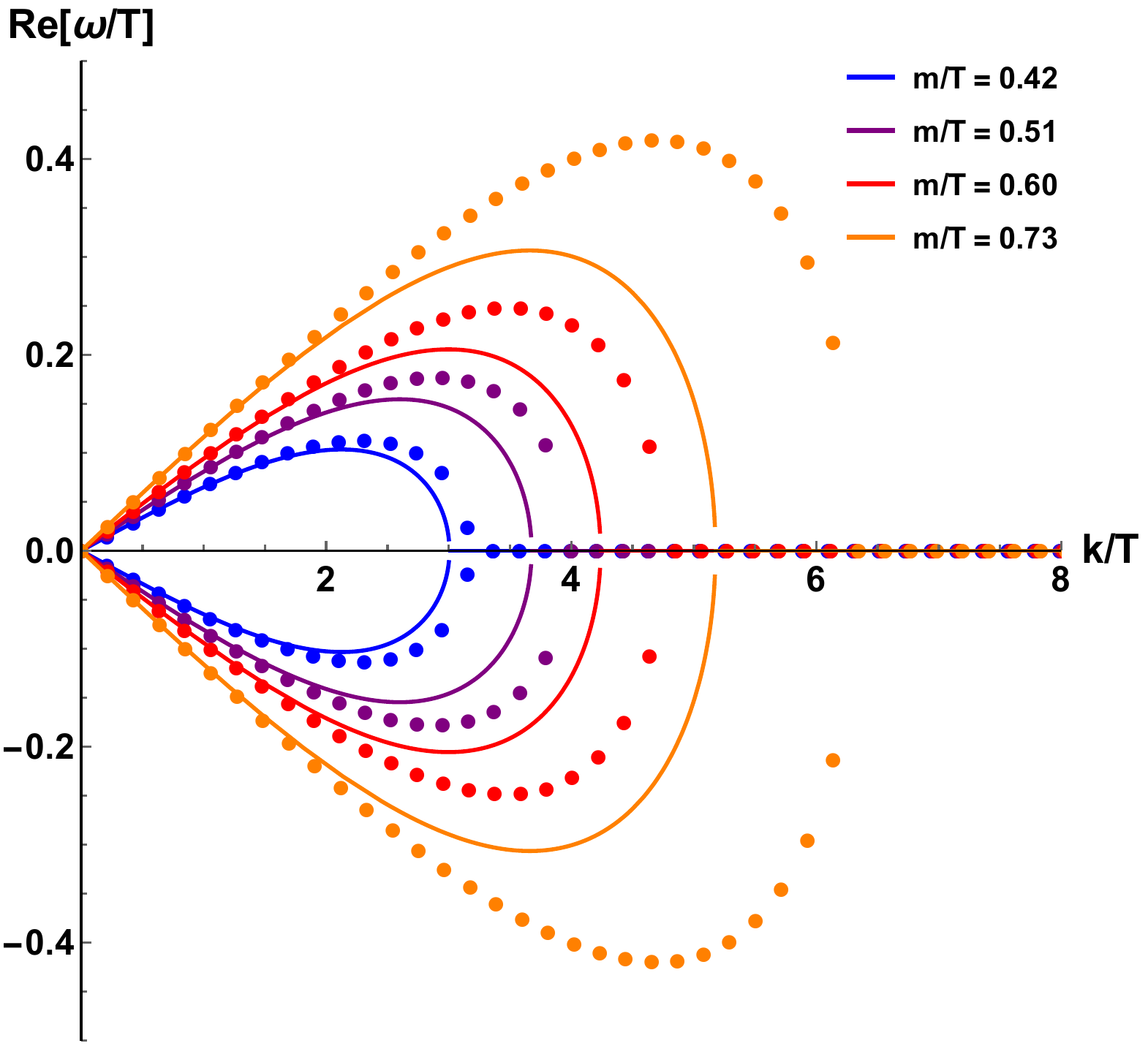}
    \caption{The transverse phonons dispersion relation. The dots are the numerical value extracted from the QNMs spectrum and the lines are the hydrodynamic formula \eqref{hydroshear}. Different colors represent different values of the dimensionless ratio $m/T$ controlling the SSB of translations. The potential is chosen as $V(X)=X^5$ but the result are analogous for all $N>5/2$.}
    \label{fig:shear}
\end{figure}

\begin{figure}
    \centering
    \includegraphics[width=7cm]{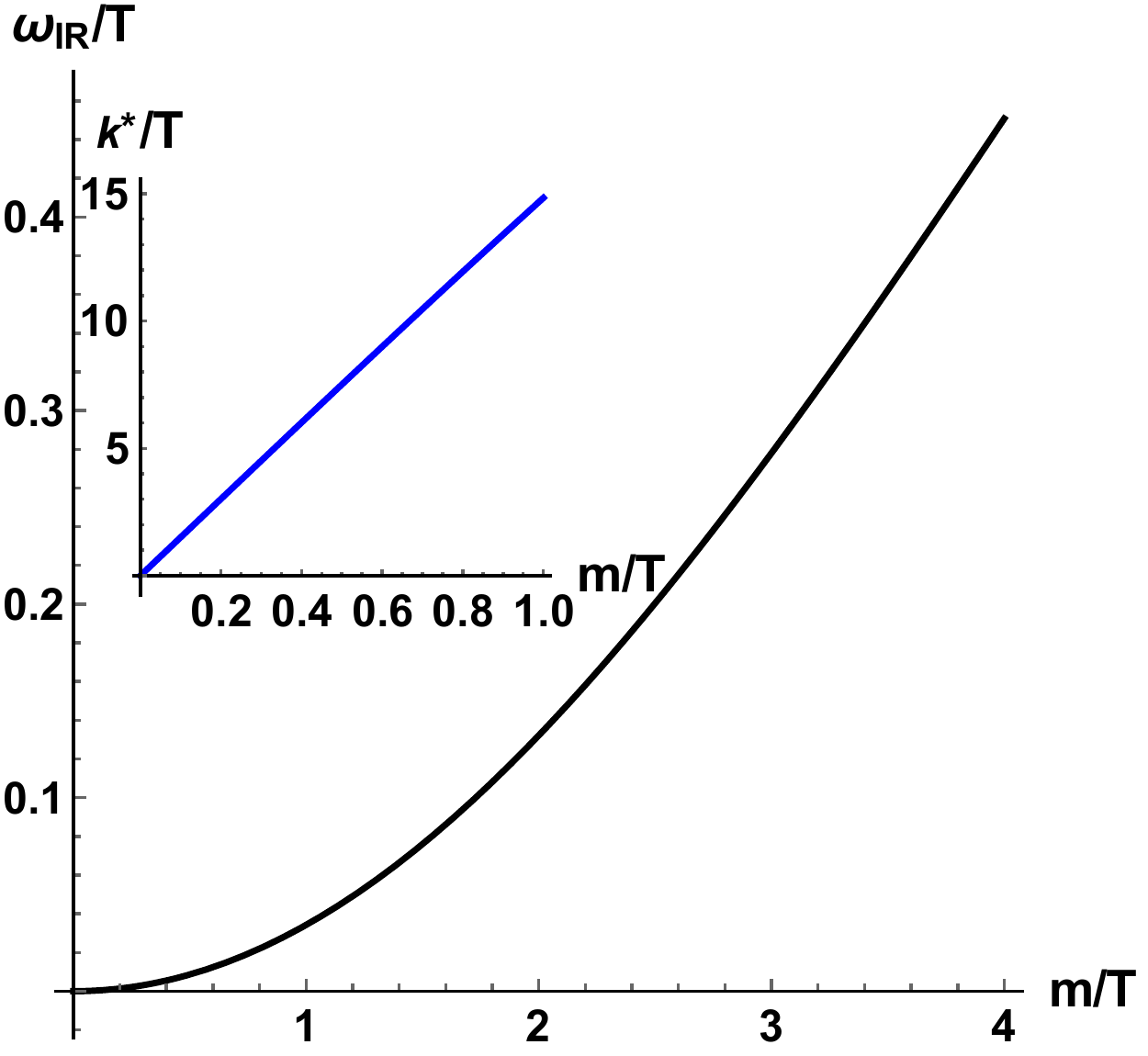}
    \quad \includegraphics[width=7.8cm]{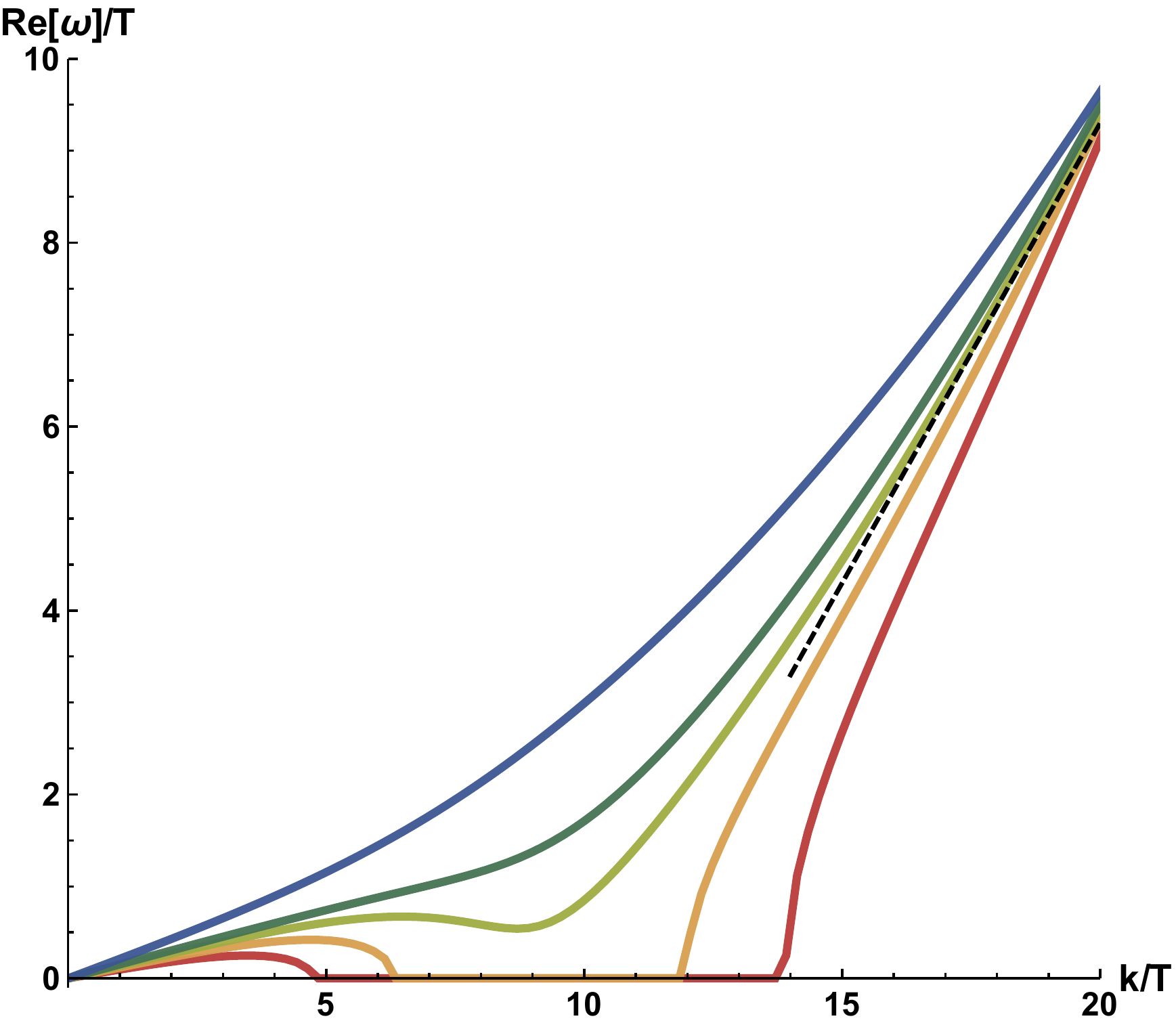}
    \caption{\textbf{Left: }The Ioffe-Regel frequency $\omega_{IR}$ from eq.\eqref{ioffe} in function of the dimensionless parameter $m/T$ for the potential $V(X)=X^5$. The inset shows the momentum $k^{\star}$ at which the real part of the dispersion relation becomes zero accordingly to the hydrodynamic formula \eqref{hydroshear}. \textbf{Right: } The dispersion relation of the transverse phonons extended until very large $k/T$ for $m/T \in [0.5,1.4]$ (from red to blue) and $V(X)=X^5$. The dashed line emphasize the $\omega=k$ asymptotic form.}
    \label{fig:ioffe}
\end{figure}

Within the choice \eqref{ch}, the parameter $m/T$ represents the dimensionless scale of SSB. The lower the temperature is, implying larger $m/T$, the stronger the SSB\footnote{Not to confuse it with the role of $m$ for the benchmark model \eqref{benchmark} used in the next sections. There the parameter $m$ represents the explicit breaking scale.}. Using Hydrodynamics \cite{PhysRevB.22.2514,Lubensky, Delacretaz:2017zxd}, we can obtain the low energy dispersion relation for the shear modes which takes the form:
\begin{equation}
    \omega_{\pm}\,=\,-\,\frac{i}{2}\,k^2\,\left(\xi\,+\,\frac{\eta}{\chi_{PP}}\right)\,\pm\,k\,\sqrt{\frac{G}{\chi_{PP}}\,-\,\frac{k^2}{4}\,\left(\frac{\eta}{\chi_{PP}}\,-\,\xi \right)^2}\,+\,\dots \, , \label{hydroshear}
\end{equation}
where $\eta$ is the shear viscosity, $G$ the shear elastic modulus, $\xi$ the Goldstone's diffusion constant and $\chi_{PP}$ the momentum susceptibility. Although equation \eqref{hydroshear} contains arbitrarily large powers of $k$, we can however only trust the terms up to $k^3$. This is because  equation \eqref{hydroshear} was derived by only including transport coefficient which are first order in derivatives. Indeed, in the low momentum limit $k/T \ll 1$, the formula \eqref{hydroshear} reduces to the expected expression for transverse and damped phonons:
\begin{equation}
    \omega\,=\,c_T\,k\,-\,i\,D_T\,k^2\,+\,\dots\, , \label{shearlow}
\end{equation}
where the transverse speed is given by $c_T^2=G/\chi_{PP}$. Equation \eqref{shearlow} was proven to hold in \cite{Alberte:2017oqx}. Notice that, at small $m/T$, the hydrodynamic formula \eqref{hydroshear} predicts that the real part of the dispersion relation of the transverse phonons closes at a critical momentum:
\begin{equation}
    k^*\,\equiv\,2\,\sqrt{\frac{G}{\chi_{PP}}\,\frac{1}{\left(\frac{\eta}{\chi_{PP}}\,-\,\xi\right)^2}}\, ,\label{closing}
\end{equation}
which is shown in fig. \ref{fig:ioffe}. Notice that the formula \eqref{closing} for $k^*$ is valid only whenever $k^*/T \ll 1$. As evident in fig. \ref{fig:shear}, when $Re(\omega(k))\,=\,0$ at large momenta, formula \eqref{closing} does not provide anymore a good approximation. This is expected since higher order transport coefficients are important for $k^*/T\sim1$.

In this paper we expand the numerical computations to large momenta, \textit{i.e.} $k/T \gg 1$. The results are shown in fig. \ref{fig:shear} for a simple choice of potential $V(X)=X^5$ and they indeed suggest that formula \eqref{hydroshear} represents a very good approximation in the hydrodynamic limit $\omega/T,k/T \ll 1$. At large momenta $k/T \gg 1$, the hydrodynamic formula is clearly not valid and higher momenta corrections make the frequency depart from the prediction in \eqref{hydroshear}. Notice that the comparison shown in fig. \ref{fig:shear} is not a fit since all the parameters in equations \eqref{hydroshear} can be computed independently. In particular we have computed:
\begin{align}
  &  G\,=\,Re\left[\mathcal{G}^R_{xyxy}\left(\omega=k=0\right)\,\right]\,,\\
  &  \eta=\,-\,\lim_{\omega\, \rightarrow 0}\left\{\frac{1}{\omega}\,Im\left[\mathcal{G}^R_{xyxy}\left(\omega,k=0\right)\,\right]\,\right\}\, ,\\
  &  \chi_{PP}\,=\,T_{tt}\,+\,T_{xx}\,=\,\frac{3}{2}\,\epsilon,\\
  &  \xi\,=\,G\,\lim_{\omega\, \rightarrow 0}\left\{\omega\,Im\left[\mathcal{G}^R_{\phi\phi}\left(\omega,k=0\right)\,\right]\,\right\} \, , \label{quantities}
\end{align}
where $\epsilon$ is the energy density of the system and $\mathcal{G}^R_{xyxy}$ and $\mathcal{G}^R_{\phi\phi}$ the Green functions of the stress tensor operator $T_{xy}$ and the scalar Goldstone $\phi$ (see appendix \ref{App1} and \ref{app2} for more details regarding the computation of these quantities).

The physics of this model is interesting because of being different from what we expect from a perfect harmonic crystal. Concretely the transverse sound is damped because of the presence of dissipative terms encoded in the $\eta,\xi$ parameters. As a consequence, there is a clear sound-diffusion crossover between a purely propagating transverse phonon at low momenta to a diffusive hydrodynamic motion at large momenta. This behaviour can be characterized using the so-called notion of Ioffe-Regel frequency (or Ioffe-Regel crossover) \cite{PhysRevB.61.12031,taraskin2002vector,beltukov2013ioffe,Shintani2008}:
\begin{equation}
    \omega_{IR}\,=\,\frac{c_T^2}{\pi\,D_T}\,=\,\frac{1}{\pi}\,\frac{G}{\eta\,+\,\xi\,\chi_{PP}} \, .\label{ioffe}
\end{equation}
This expression estimates the frequency at which the phonons stop to propagate and start behaving in a  diffusive or quasi-localized way. In our model the Ioffe-Regel frequency $\omega_{IR}$ grows with the dimensionless ratio $m/T$ as shown in fig. \ref{fig:ioffe}. This is in qualitative agreement with the behaviour of $k^{\star}$ (see fig. \ref{fig:ioffe}) and can be simply understood noticing that at small temperature the dissipative effects become weaker and therefore the phonons propagate longer. This behaviour suggests that for larger and larger $m/T$ the system behaves more and more as a perfect crystal. Dissipation decreases increasing $m/T$ and the phonons become less and less damped.

The Ioffe-Regel crossover is a typical feature of disordered and viscoelastic materials such as amorphous solids and glasses. The competition between a propagative and a damping term in the dispersion relation of the phonons can be moreover a possible explanation for the universal presence of the Boson Peak anomaly in glasses and ordered crystals \cite{baggioli2018universal,baggioli2018soft}. Additionally, it represents another confirmation regarding the nature of the sometimes denoted holographic ``homogeneous'' models for the breaking of translational invariance. The models at hand are clearly different from an ordered crystal since they miss any notion of a preferred wave vector, Brillouin zone or lattice cutoff. Nevertheless, their physics appears, at least qualitatively, consistent with what known for amorphous solids where translational invariance is broken but no long-range order is present.
\\
If we take large values of the parameter $m/T$ and we extend the computation to very large momenta $k/T \gg 1$ we can notice some important features. Decreasing the value of temperature the phonon stops to propagate at larger and larger frequencies/momenta. Interestingly, when $m/T \gg 1$, the real part of the dispersion relation does not close anymore and the phonon propagates all the way up to infinite momentum\footnote{This feature is obviously non realistic and is related to the absence of any UV cutoff in the holographic theory. In a realistic material, where a natural UV cutoff is provided by the lattice spacing $a$, this does not happen. The speed of sound becomes zero at the edge of the Brillouin zone and the phonons stop to propagate.}. In the intermediate regime, a local minimum, reminiscent of a superfluid roton excitation, appears in the dispersion relation around $k/T \sim 10$. On the contrary for very large $m/T$ the phonon with initial speed $c_T^2=G/\chi_{PP}$ connects directly with the large momenta regime $\omega=k$. In that high energetic regime the physics is completely dominated by the conformal UV fixed point and all the dispersion relations asymptotes the form $\omega= k$ at large momenta $k/T \gg 1$. Finally, notice that this feature also happens in the cases where the phonons stop to propagate at a certain momentum $k_{IR}$. For those values the transverse mode real part is zero only in an interval of momenta $k \in [k_{IR},k_{UV}]$. Clearly, all these features cannot be captured by hydrodynamics. It would be very interesting to understand them better with possible alternative methods such as the resummation of the hydro expansion like done in \cite{Baggioli:2018bfa,Withers:2018srf}. Interestingly, dispersion relations with wavevector cutoffs have recently raised lot of interest in the discussion of liquid and solid phases of matter \cite{Baggioli:2018vfc,Grozdanov:2018fic,Baggioli:2019jcm}.
\section{About the absence of ``standard'' phase relaxation}\label{sec: phase}
The interplay of explicit and spontaneous breaking of translational invariance has been advocated to be a possible explanation for the optical conductivity of bad metals \cite{Delacretaz:2016ivq}. In that context, a very important role is played by \textit{phase relaxation} which is usually denoted with the symbol $\Omega$. Phase relaxation is the tendency of the material to restore the disordered phase via the relaxation of the phase of the Goldstone boson. It is analogous to the phase relaxation mechanism in superconductors \cite{Davison:2016hno} and it can be mediated by the presence of topological defects in the material known as dislocations  \cite{Beekman:2016szb,Delacretaz:2017zxd}. In this case, when $\Omega$ is driven by dislocations, it does not depend on the EXB

Recently, two works \cite{Amoretti:2018tzw,Andrade:2018gqk} discussed the possible presence of \textit{phase relaxation} and \textit{fluctuating order} in the context of holography. In this small section we aim to clarify the role of phase relaxation in the class of homogeneous models considered. Moreover, we broaden the definition of phase relaxation. In fact, as we will see, we can rule out any mechanism independent of the explicit breaking which enters into hydrodynamics in the same way as phase relaxation $\Omega$. In particular, we suggest that no phase relaxation nor fluctuating order \footnote{By ``fluctuating  order'' we mean the presence of a non-constant (in space or time) order parameter. As explained in detail in \cite{kivelson210683detect}, the definition, and in particular the experimental observation, of fluctuating order is not a simple task.}  stricto sensu are present, so far, in any of the bottom-up holographic models discussed in the literature.

This idea was already present in \cite{Andrade:2018gqk}; here we give more evidence for the cause.
In order to do that, let us focus first on the shear correlator $\langle T_{xy} T_{xy}\rangle$ and in particular on the frequency dependent viscosity $\eta(\omega)$:
\begin{equation}
    \eta(\omega)\,\equiv\,\frac{1}{i\,\omega}\,\mathcal{G}^R_{T_{xy}T_{xy
}}\left(\omega,k=0\right) \, .\label{viscvisc}
\end{equation}
In presence of phase relaxation $\Omega$, and in absence of explicit breaking, the low frequency dependence of the viscosity is expected to be \cite{Delacretaz:2017zxd} :
\begin{equation}
    \eta(\omega)\,=\,\frac{G}{\Omega\,-\,i\,\omega}\,+\,\eta\,+\,\dots\, ,\label{nono}
\end{equation}
where $G$ and $\eta$ are the shear elastic modulus and the shear viscosity. In other words the phase relaxation parameter induces a ``Drude peak'' into the frequency dependent viscosity.\\

\begin{figure}
    \centering
    \includegraphics[width=7.5cm]{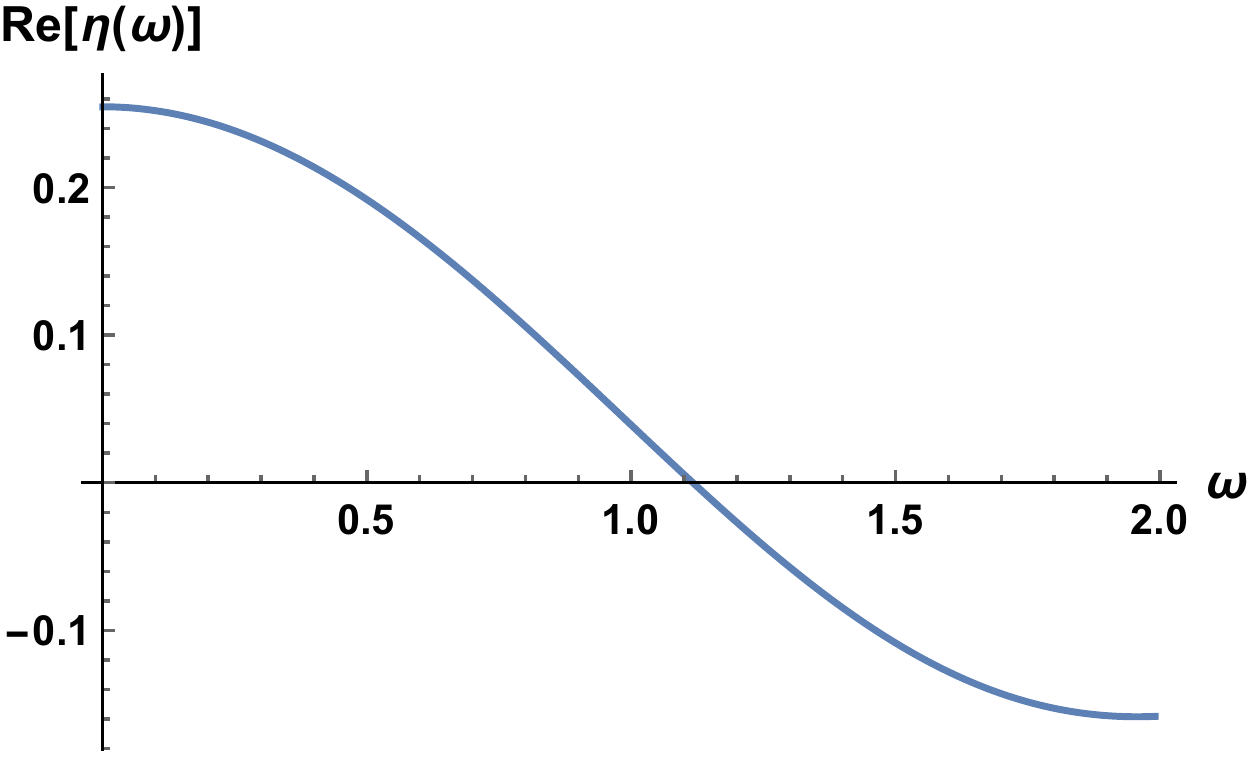}
    \quad 
     \includegraphics[width=7.5cm]{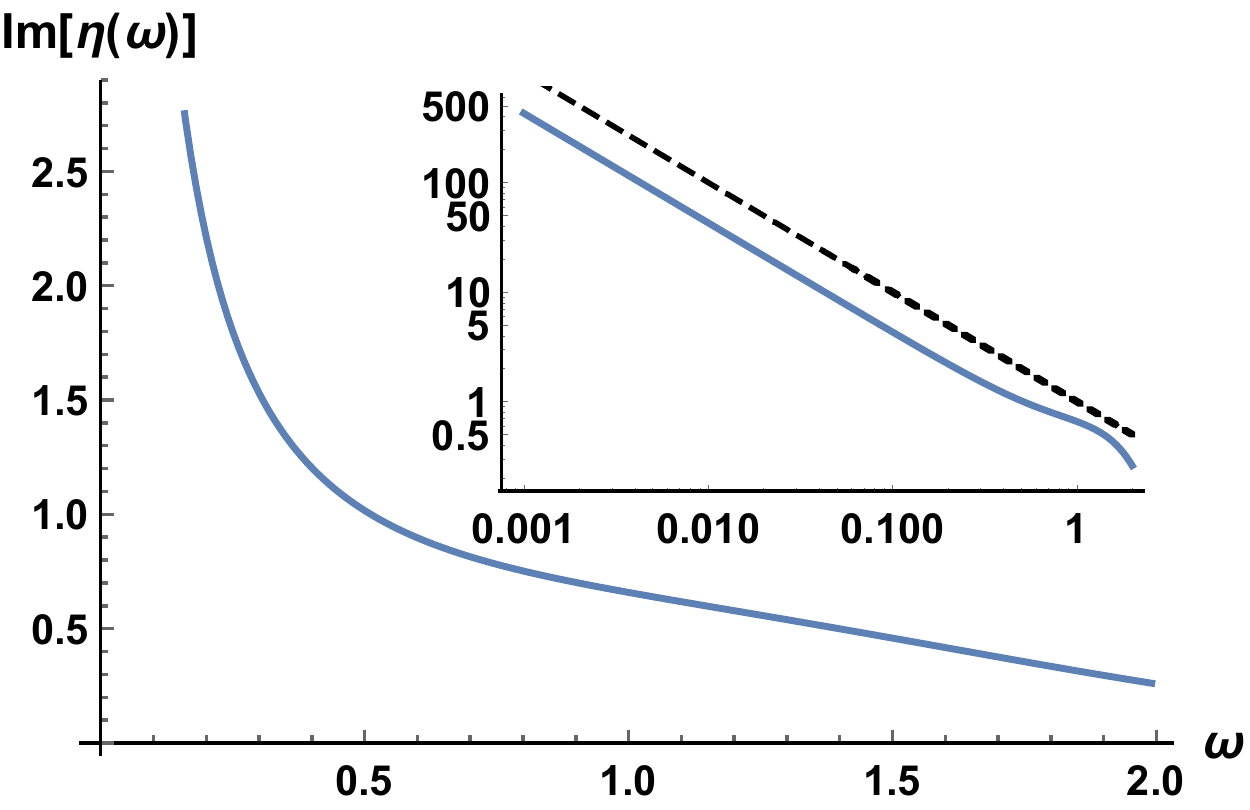}
    \caption{The frequency dependent viscosity \eqref{viscvisc} for the potential $V(X)=X^3$ at $m/T \approx 3.5$. \textbf{Left: }The real part showing the zero frequency value $\eta$. \textbf{Right: } The imaginary part showing the $1/\omega$ pole and the absence of $\Omega$. The inset is a log-log plot to show more clearly the low frequency behaviour.}
    \label{fignophase}
\end{figure}

As a matter of fact, it is immediate to check the presence of phase relaxation computing the shear correlator from the equation for the shear fluctuations\footnote{This result applies to systems way more general than ours. In particular, given a specific holographic model, with a radial dependent graviton mass $\mathcal{M}(u)$ (which in isotropic systems can be derived directly from the stress tensor),  it was proven in \cite{Hartnoll:2016tri} that the equation for the shear mode is:
\begin{equation}
    \Box H\,=\,\mathcal{M}^2(u)\,H \, ,\label{gengen}
\end{equation}
where $\Box$ is the Laplacian operator on the background metric. Starting from the shear equation \eqref{gengen} we have not found evidence for the appearance of a Drude pole and therefore of a proper phase relaxation rate $\Omega$.}:
\begin{equation}
    H''\,+\,H'\,\left(-\frac{2}{u}\,+\,\frac{2\,i\,\omega}{f}\,+\,\frac{f'}{f}\right)\,+\,H\,\left(-\,\frac{2\,i\,\omega}{u\,f}\,-\,\frac{2\,m^2\,V_X}{f}\right)\,=\,0\, , \label{eqshear}
\end{equation}
where $H(u)\equiv \delta g^x_y(u)$. The Green functions can be extracted using the standard holographic dictionary (for more details see \cite{Alberte:2016xja}).

Solving equation \eqref{eqshear} numerically for various potentials, we can immediately test if the phase relaxation parameter $\Omega$ is present in our model or not. The generic outcome is that there is no Drude peak in the frequency dependent viscosity \eqref{viscvisc}. An example of the results is shown in fig. \ref{fignophase}.

Consequently, we can affirm that no phase relaxation mechanism as incorporated in the hydrodynamics of \cite{Delacretaz:2017zxd} is present in our model.

\begin{figure}[hbtp]
    \centering
    \includegraphics[width=7.5cm]{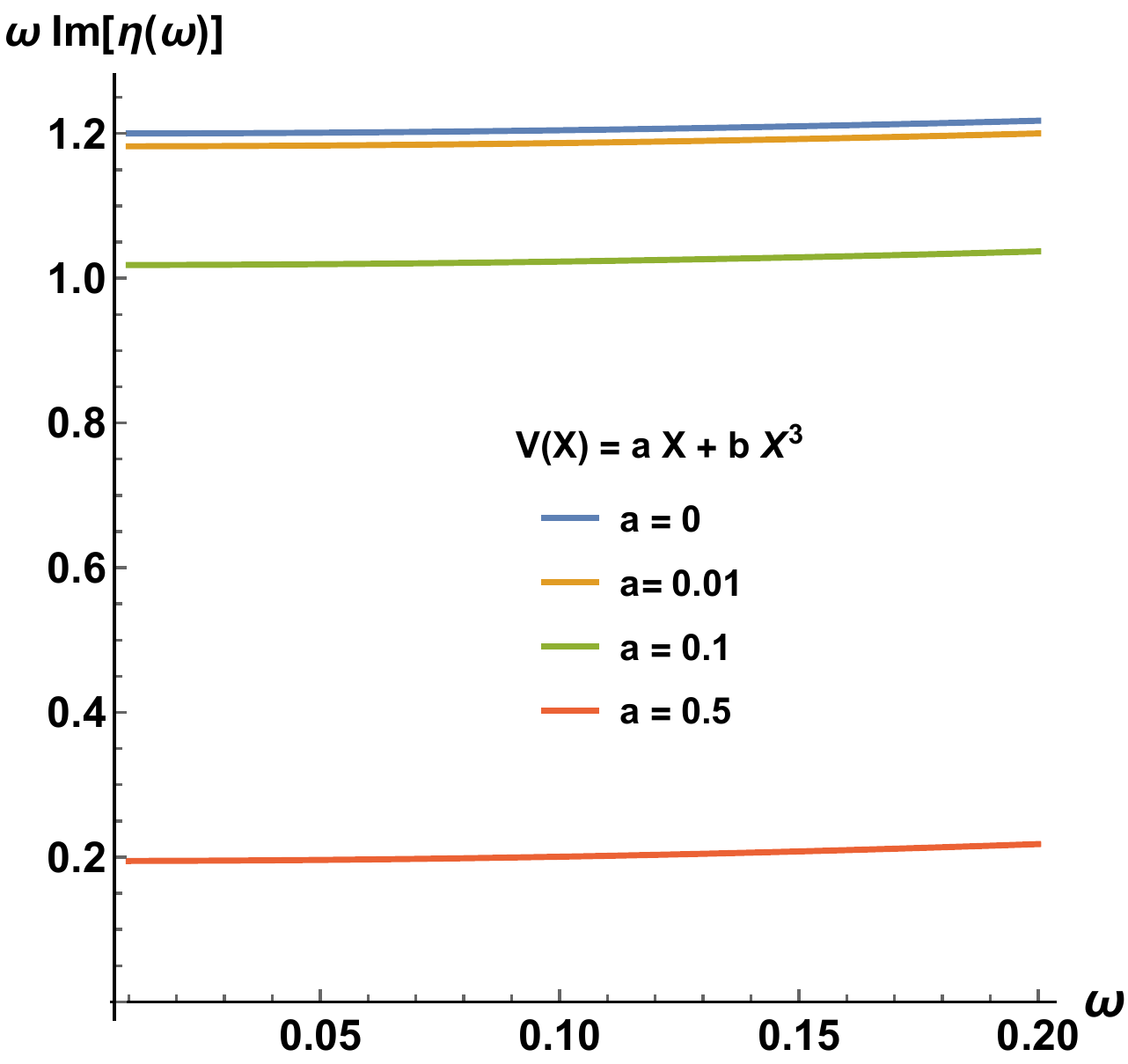}
    \caption{The imaginary part of the frequency dependent viscosity \eqref{nono} for a potential of the type $V(X)= a\,X\,+\,b\,X^3$.  We see the presence of a $\sim 1/\omega$ pole both in the purely spontaneous and pseudo-spontaneous regime.}
    \label{proof}
\end{figure}

In the following, however, we will discuss a term, playing an analogous role of the phase relaxation $\Omega$ in the optical conductivity, is present in our model as well as in the already cited works \cite{Amoretti:2018tzw,Andrade:2018gqk}. This term, denoted in our manuscript as $\bar{\Omega}$, is absent in the hydrodynamic description of \cite{Delacretaz:2017zxd} and its nature is not yet well understood. In particular, it is not clear if the $\bar{\Omega}$ term can be thought of as a purely additional contribution to the phase relaxation $\Omega$. Indeed, even in the pseudo-spontaneous case where $\bar{\Omega}\neq 0$, we do not find any Drude structure in the frequency dependent viscosity \eqref{nono}, as shown in figure \ref{proof}.
\\
This result does not contradict the results of \cite{Delacretaz:2017zxd} because the addition of explicit breaking, even if small, invalids the assumption which leads to formula \eqref{nono}. A generalization of that formula in presence of explicit breaking is necessary to match our holographic results with the hydrodynamic framework. Work in this direction is on-going.
\section{AC electric conductivity}\label{sec: ac}
In this section we compute the electric AC conductivity $\sigma(\omega)$ in our model. We obtain the conductivity from the current-current Green function:
\begin{equation}
    \sigma(\omega)\,=\,\,\frac{1}{i\,\omega}\,\mathcal{G}_{JJ}(\omega)\, ,
\end{equation}
at zero momentum $k=0$. 

In the purely spontaneous case, $V(X)=X^N$ with $N>5/2$ or $\beta=\infty$ for the potential \eqref{benchmark}, the conductivity was computed in \cite{Alberte:2017oqx} and it takes the expected form (at low frequency):
\begin{equation}
    \sigma(\omega)\,=\,\sigma_{inc}\,+\,\frac{\rho^2}{\chi_{PP}}\,\left(\frac{i}{\omega}\,+\,\delta(\omega)\,\right)\,+\,\dots\, ,
\end{equation}
where $\sigma_{inc}$ is the so-called incoherent conductivity \cite{Davison:2015taa}.

Oppositely, we can consider no spontaneous breaking and the limit of small explicit breaking which leads to a large relaxation time for the momentum operator:
\begin{equation}
    \frac{1}{\tau\,T}\,=\,\frac{\Gamma}{T}\,\ll\,1\, .
\end{equation}
In absence of spontaneous breaking the conductivity would be well fitted by a Drude form:
\begin{equation}
    \sigma(\omega)\,=\,\sigma_0\,+\,\frac{\rho^2}{\chi_{PP}}\,\frac{1}{\Gamma\,-\,i\,\omega}\, ,
\end{equation}
which is typical of systems where momentum is an almost conserved quantity \cite{Vegh:2013sk}.

\begin{figure}[hbt]
    \centering
    \includegraphics[width=7.5cm]{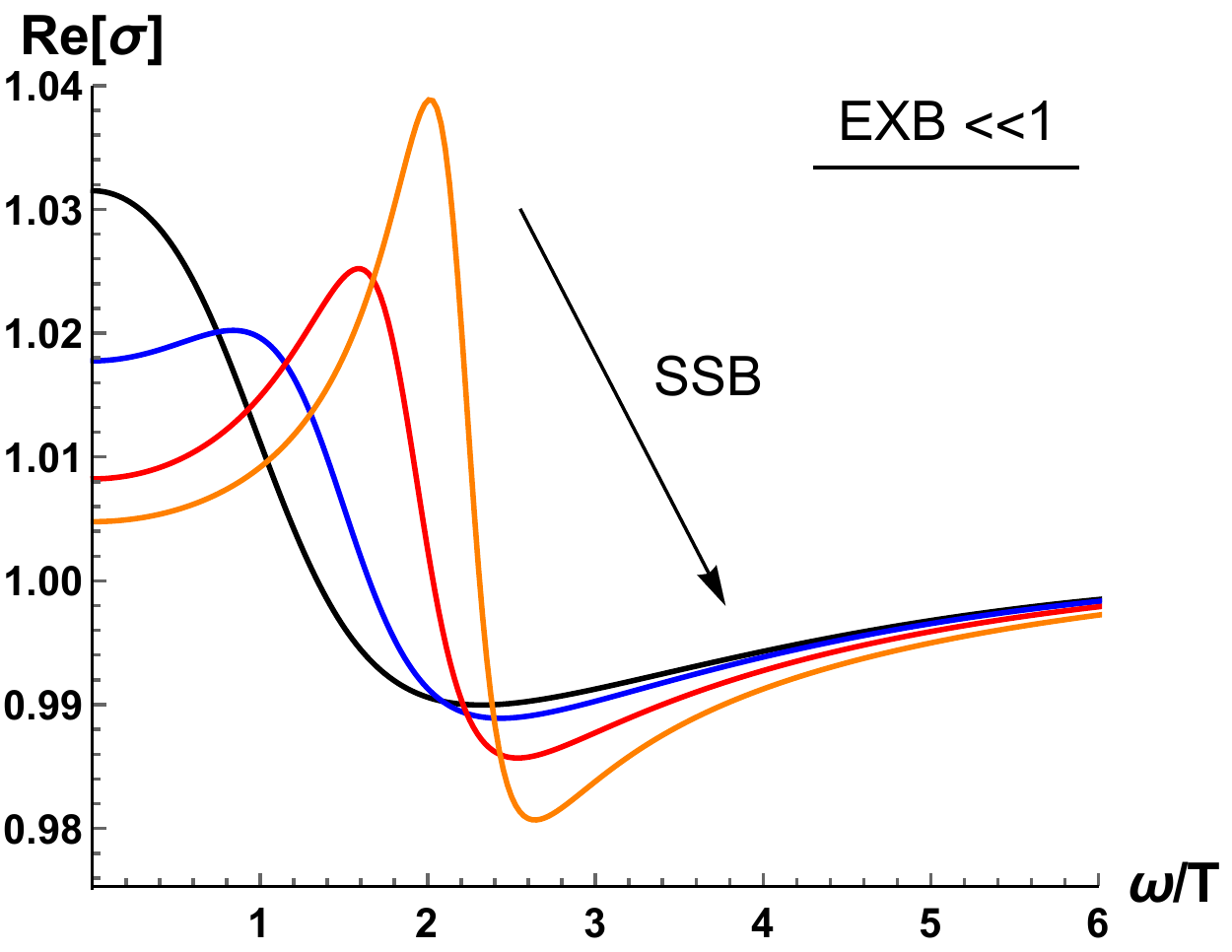}
    \quad
    \includegraphics[width=7.5cm]{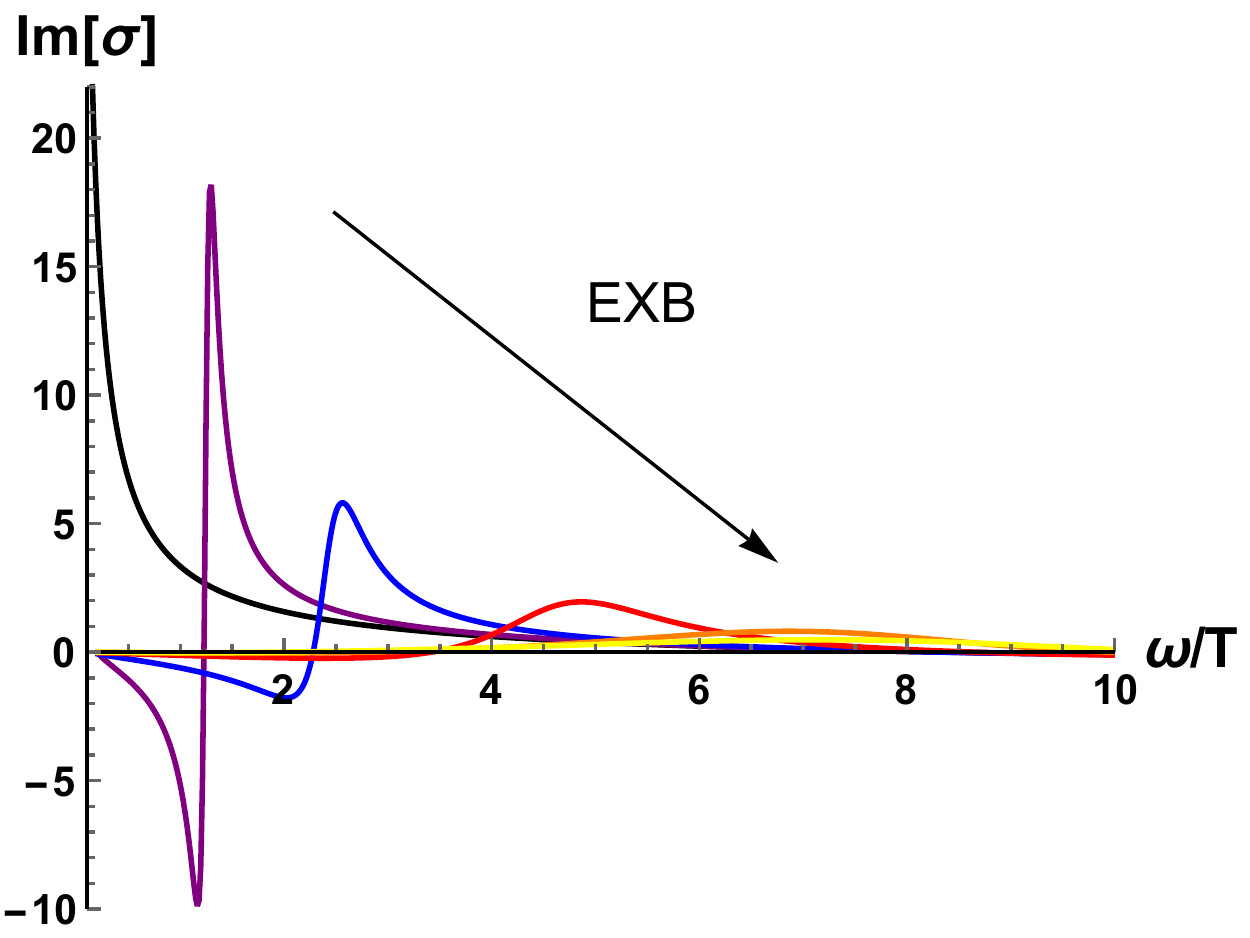}
    \caption{\textbf{Left: }Real part of the electric conductivity increasing the strength of SSB starting from a situation with a small but finite EXB. From black to orange the breaking becomes more and more spontaneous. We fixed $T/m=1$ and $\mu/T=0.5$. From black to orange  increasing $\beta \in [1,10]$. The transition from a Drude peak to a coherent pinned response is evident. \textbf{Right: }Imaginary part of the conductivity starting from a pure SSB situation (black line) and adding EXB. This plot is done using the potential $V(X)=\mathfrak{A}X+\mathfrak{B}X^5$. The transition is now from a $i/\omega$ pole to a pinned response going at larger and larger frequencies and being more and more damped until disappearing.}
    \label{AC1}
\end{figure}

More interesting is the interplay of the explicit and spontaneous breaking, in the limit where the explicit scale is much smaller than the spontaneous one, \textit{i.e.} the pseudo-spontaneous limit. Taking this assumption, the explicit breaking is just a small deformation to the SSB pattern such that a gapped but light mode can be still identified in the spectrum. Given the interplay between spontaneous and explicit symmetry breaking, the singularities of the electric current two point function can be now found from solving the simple expression:
\begin{equation}
   \left(\bar{\Omega}-\,i\,\omega\right)\,\left(\Gamma\,-\,i\,\omega\right)\,+\,\omega_0^2\,=\,0\, ,
\end{equation}
where $\omega_0$ is the so-called pinned frequency or in other words the mass gap of the pseudo-goldstone boson. As discussed previously, we do not find traces of phase relaxation in the shear viscosity \ref{nono}. Nevertheless the structure of the poles in the correlators and the form of the optical conductivity are consistent with an additional parameter which, in presence of $\Gamma$ and $\omega_0$, enters only in certain transport properties as the phase relaxation term $\Omega$ would. The effects of the interplay between the two mechanism of symmetry breaking and in particular of the presence of a finite pinning frequency $\omega_0$ will be evident in the AC conductivity as already shown in \cite{Baggioli:2014roa}. The spectral weight is going to be transferred from the Drude peak to an intermediate frequency scale which is set by the pinning frequency $\omega_0$.

\begin{figure}
    \centering
    \includegraphics[width=7.5cm]{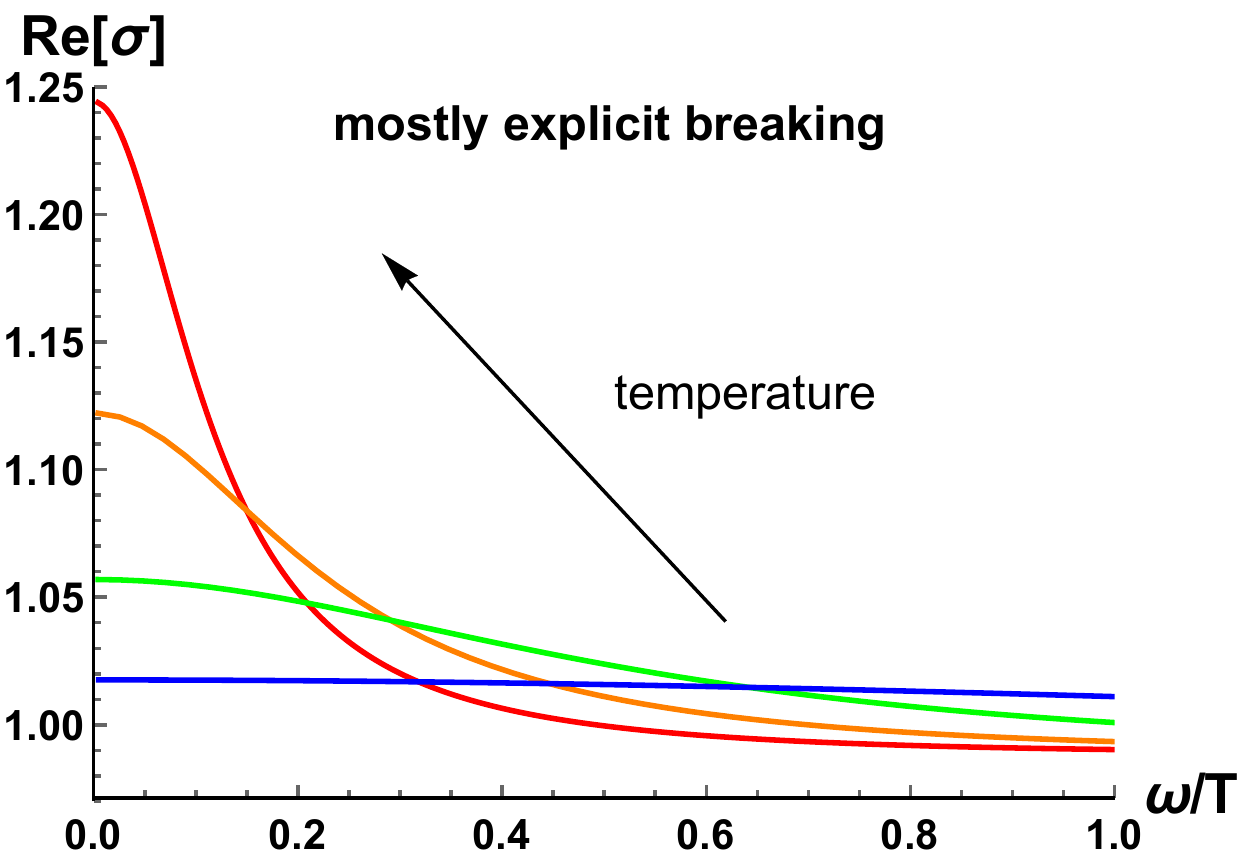}
    \quad
    \includegraphics[width=7.5cm]{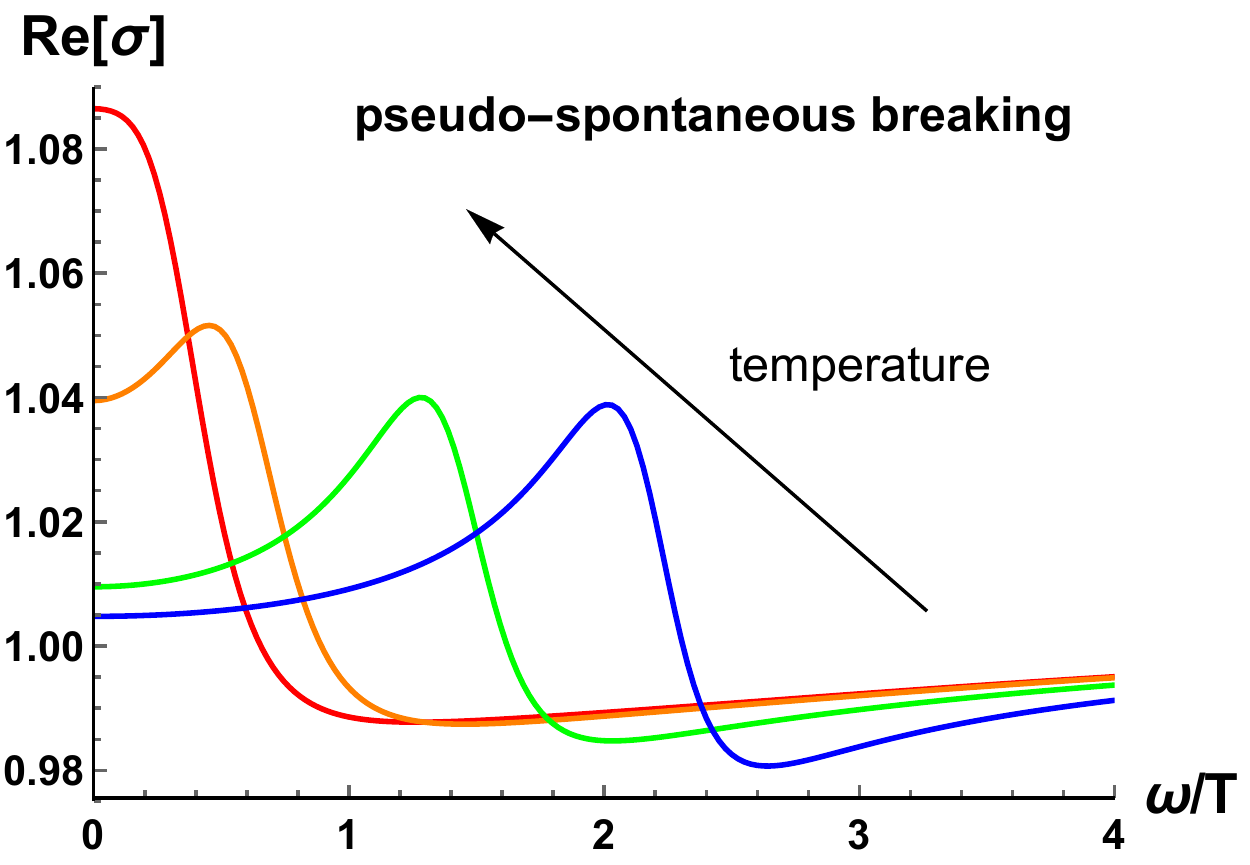}
    \caption{AC conductivity moving the dimensionless temperature parameter at fixed chemical potential $\mu/T=0.5$. \textbf{Left: }A mostly explicit case $\beta=0.1$. The behaviour is a coherent-incoherent crossover. \textbf{Right: }A mostly spontaneous case $\beta=10$. Now the behaviour is from a Drude peak to a coherent pinned response.}
    \label{AC2}
\end{figure}

We compute the AC conductivity numerically (for more details see Appendix \ref{App1}) and display the results in figs. \ref{AC1} and \ref{AC2}. In particular, in fig. \ref{AC1} we dial the ratio between the spontaneous and the explicit breaking keeping fixed the other parameters. For $\beta$ small, the breaking is mostly explicit and the AC conductivity shows a Drude peak form; on the contrary, when increasing $\beta$ (i.e. when the breaking pattern becomes more and more spontaneous), the spectral weight at zero frequency is transferred to finite frequency producing the typical AC form of a pinned system. Moreover, increasing the spontaneous breaking, the peak moves to higher frequency in accordance to the GMOR relation \cite{PhysRev.175.2195} and its width becomes smaller.

Moreover, we can start from the purely SSB case and add more and more EXB. In such case (see right panel of fig. \ref{AC1}), the $i/\omega$ pole moves to finite frequency and becomes more and more damped till disappearing for large EXB. This picture is consistent with what observed in the QNMs spectrum of the system (see next section).

In fig. \ref{AC2} we keep fixed the parameter $\beta$ and we move the dimensionless parameter $m/T$. At low $\beta$ the breaking is mainly explicit and moving the dimensionless temperature the AC conductivity shows the crossover between a coherent regime with a sharp Drude peak to an incoherent regime with a flat response \cite{Davison:2015taa}. The crossover is produced by a collision between the two lowest poles, which appears very far from the hydrodynamic limit, \textit{i.e.} $\textrm{Im}(\omega)/T \sim 8$. The collision produces a couple of poles with a real part but their signature on the AC conductivity is smoothed out since they are highly damped. In contrast to that, at large $\beta$, the breaking is pseudo-spontaneous and the collision happens within the hydrodynamic limit or in other words at small frequencies. As a consequence, the system shows a crossover from a sharp Drude peak to a pinned regime with a coherent peak at low frequency. 

\begin{figure}[hbtp]
    \centering
    \includegraphics[width=7.5cm]{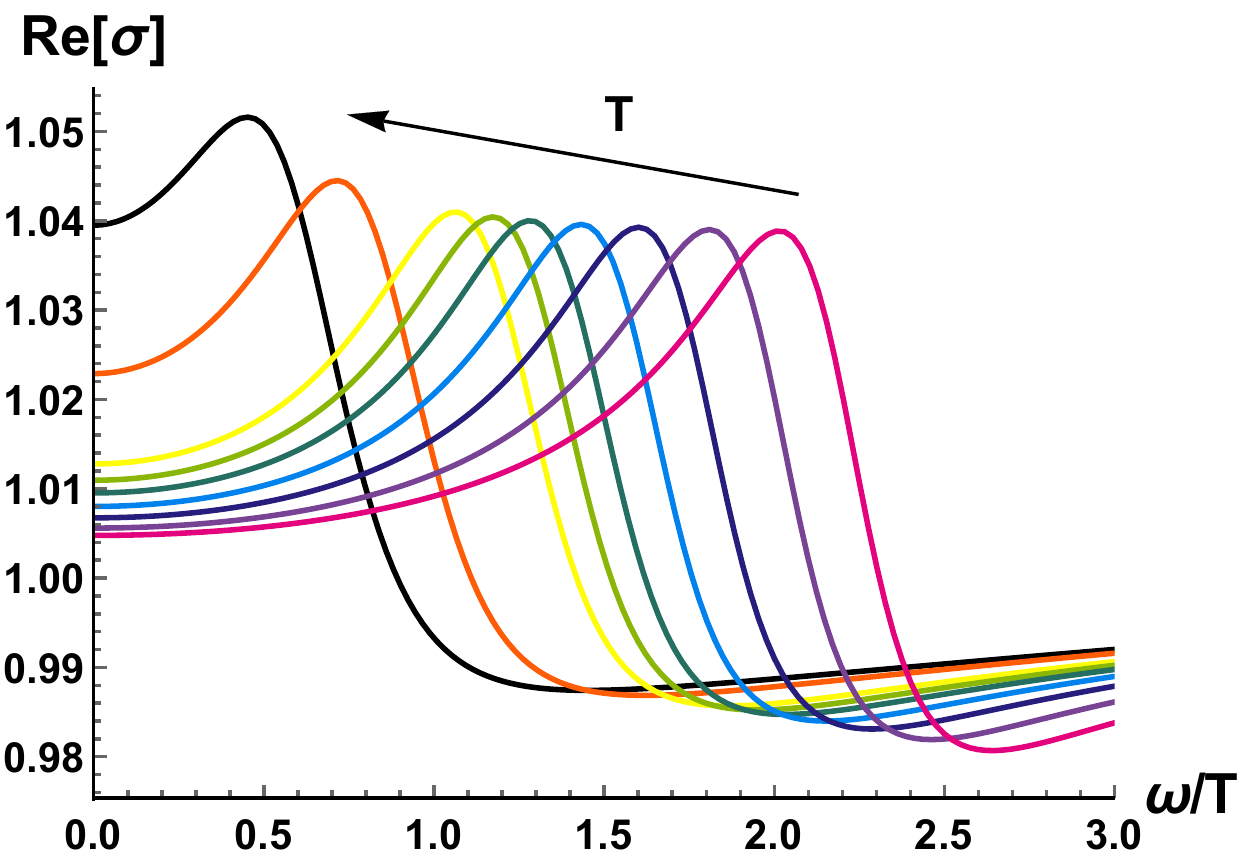}
    \quad
    \includegraphics[width=7.5cm]{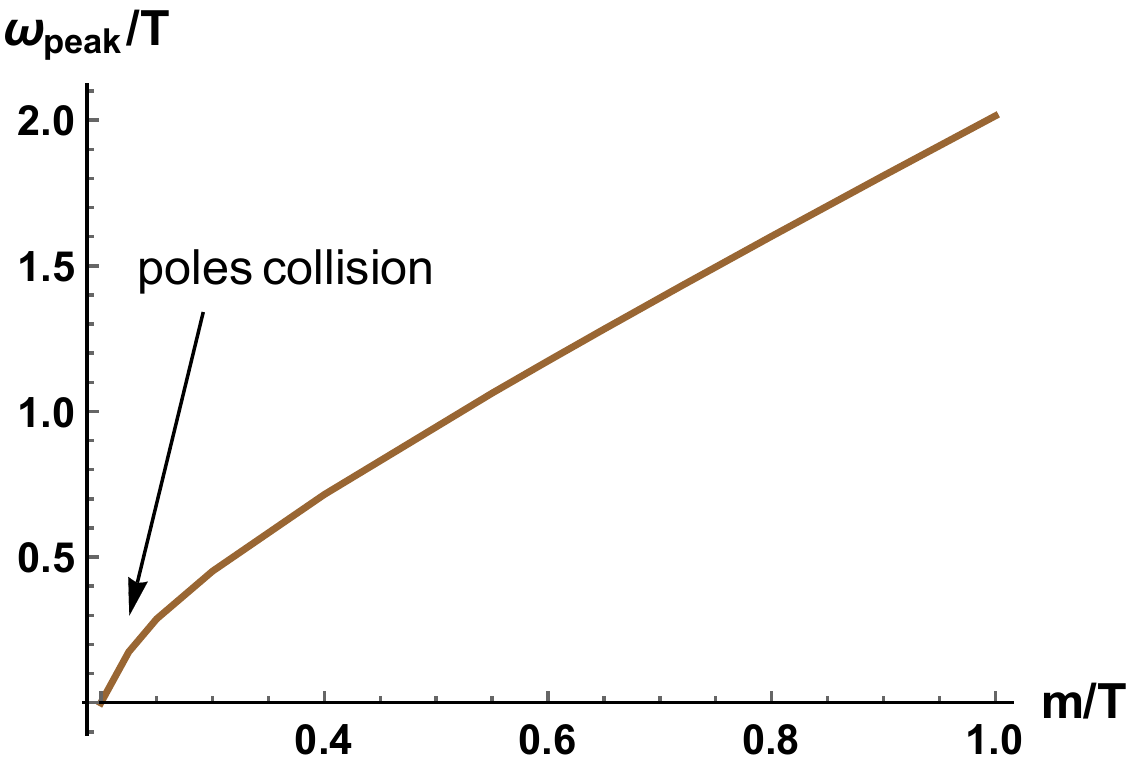}
    \caption{The behaviour of the off-axes peak in function of the dimensionless temperature. We fixed $\beta=10$ and we dial $m/T \in [0.2,1]$.}
    \label{ins}
\end{figure}

Let us also emphasize that in the pseudo-spontaneous limit the system displays an insulating behaviour with $d\sigma /dT >0$ and that the pinned peak moves at higher frequencies decreasing the temperature, see fig. \ref{ins}. The off-axes peak develops after the poles collision and it moves in a square root fashion to higher energies decreasing the dimensionless temperature. The dynamics displays an insulating nature, in contrast with what was obtained in \cite{Delacretaz:2016ivq} and what was found in \cite{Amoretti:2017axe}. In that case, the presence of a bulk dilaton field with a specific potential was needed to have the metallic behaviour at low temperature. Generally, a pinned response always corresponds to an insulating state at low temperature\footnote{A simple argument to follow is just the sum rule for the optical conductivity.}. This is the usual case for charge density waves \cite{RevModPhys.60.1129} and for the other existing holographic models \cite{Baggioli:2014roa,Andrade:2017cnc,Andrade:2018gqk}.

Finally, let us notice that the square root behaviour of the peak in function of the dimensionless parameter $m/T$ is already an hint towards the validity of the GMOR relation \cite{PhysRev.175.2195} which we will analyze in detail in the following section.
\section{Explicit VS spontaneous breaking and the collective modes}\label{sec: pseudo}
In this section we consider the interplay of spontaneous and explicit breaking of translations using the potential:
\begin{equation}
    V(X)\,=\,X\,+\,\beta\,X^N\,,\quad N\,>\,5/2\, ,
\end{equation}
and for most of the discussion $N=5$.

The scale of explicit breaking in the model is represented by the graviton mass in the UV (at the AdS boundary $u=0$) that reads
\begin{equation}
  \boxed{  \textit{EXPLICIT SCALE}\, \equiv \langle EXB \rangle \,=\, m_{UV}\,=\,m\, ,}
\end{equation}
and it coincides with (the square root of) the coefficient of the linear term in the potential $V(X)$. As evidence for this, we can observe that the  momentum relaxation rate is proportional to such a scale and in particular for small explicit breaking (at leading order in $m/T$) reads \cite{Davison:2013jba}:
\begin{equation}
  \Gamma\,\propto \langle EXB \rangle^2\,\quad \Rightarrow \quad \Gamma\,\propto\,m^2 \, , \label{expscale}
\end{equation}
which is in agreement with the results obtained using the Memory Matrix formalism \cite{Hartnoll:2012rj,Lucas:2015vna}. In the limit of zero explicit breaking, i.e. $m=0$, the relaxation time for the momentum operator is infinite and the DC conductivity diverges.

On the contrary, we expect the spontaneous scale to be proportional to the coefficient of the $X^N$ term appearing in the potential $V(X)$. By analogy, we can identify the SSB scale as:
\begin{equation}
    \boxed{\textit{SPONTANEOUS SCALE}\, \equiv \langle SSB \rangle \,\propto\,m\,\sqrt{\beta}\, . } \label{spontscale}
\end{equation}
This assumption is corroborated by the definition of the elastic modulus $G$ (which is meaningful only in the limit of mostly spontaneous breaking). In particular, in the limit of small explicit breaking $m/T \ll 1$ and  $\beta \gg 1$, we have \cite{Alberte:2016xja}:
\begin{equation}
    G\,\sim\,m^2\,\beta\,\sim\,\langle SSB \rangle ^2 \, ,
\end{equation}
which is the expected dependence. It is interesting to notice that indeed the competition between EXB and SSB can be physically identified with the competition between the momentum dissipation rate $\Gamma$ and the elastic modulus $G$. As already discussed, in physical terms, the explicit breaking parametrizes the dissipation of momentum while the spontaneous breaking the elastic, and non dissipative, properties of the materials.

Given these definitions, we can see that the ideal indicator to determine if the breaking is mostly explicit or spontaneous is the dimensionless ratio of the two scales:
\begin{equation}
  \boxed{  \frac{\langle EXB \rangle}{\langle SSB \rangle}\,=\,\frac{1}{\sqrt{\beta}} \, .}
\end{equation}
This amounts to say that at $\beta=0$ the breaking is totally explicit, while at $\beta=\infty$ the breaking is totally spontaneous. We define the pseudo-spontaneous case as the limit where:
\begin{equation}
   \boxed{\textit{PSEUDO-SPONTANEOUS:}\quad \frac{\langle EXB \rangle}{\langle SSB \rangle}\,\ll 1\,\Longrightarrow\quad \,\beta\,\gg \,1 \, .}\label{pseudodef}
\end{equation}

\begin{figure}[hbt]
    \centering
    \includegraphics[width=4.8cm]{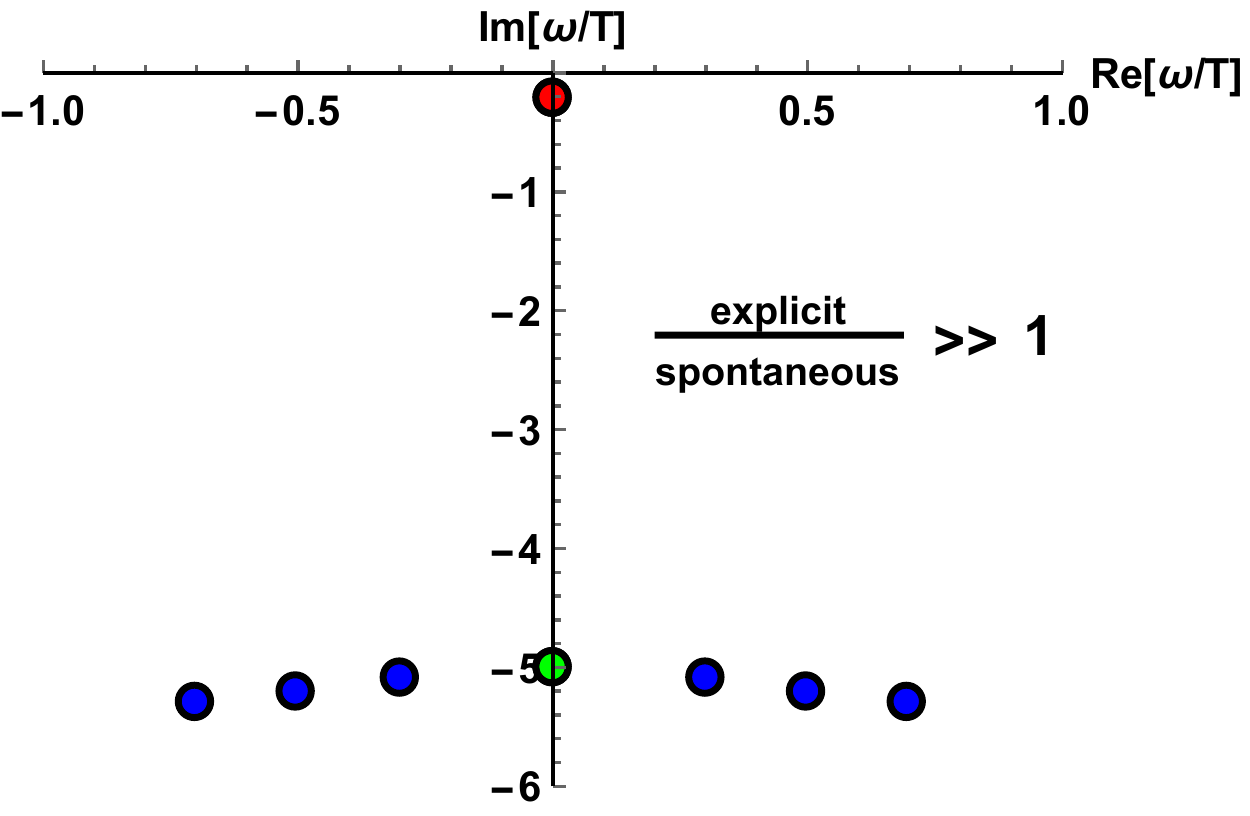}
    \quad
     \includegraphics[width=4.8cm]{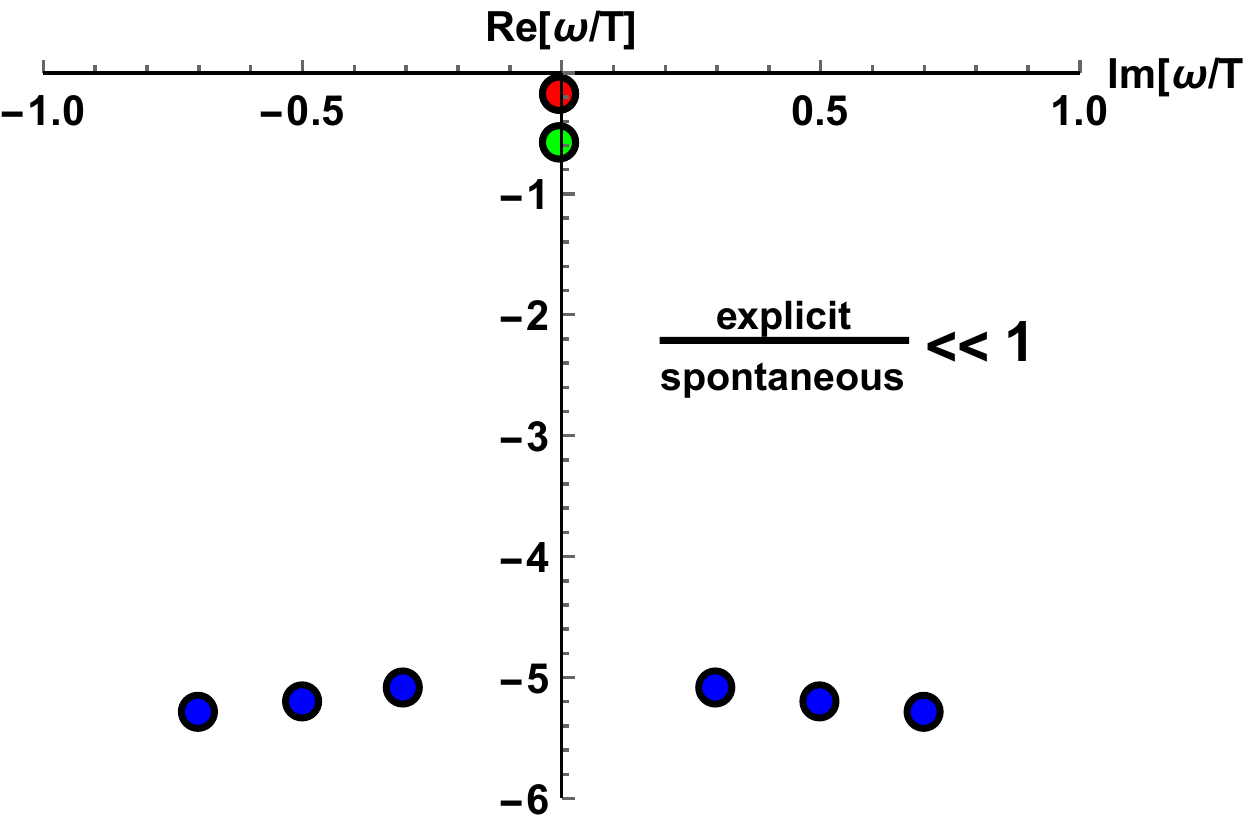}
     \quad
      \includegraphics[width=4.8cm]{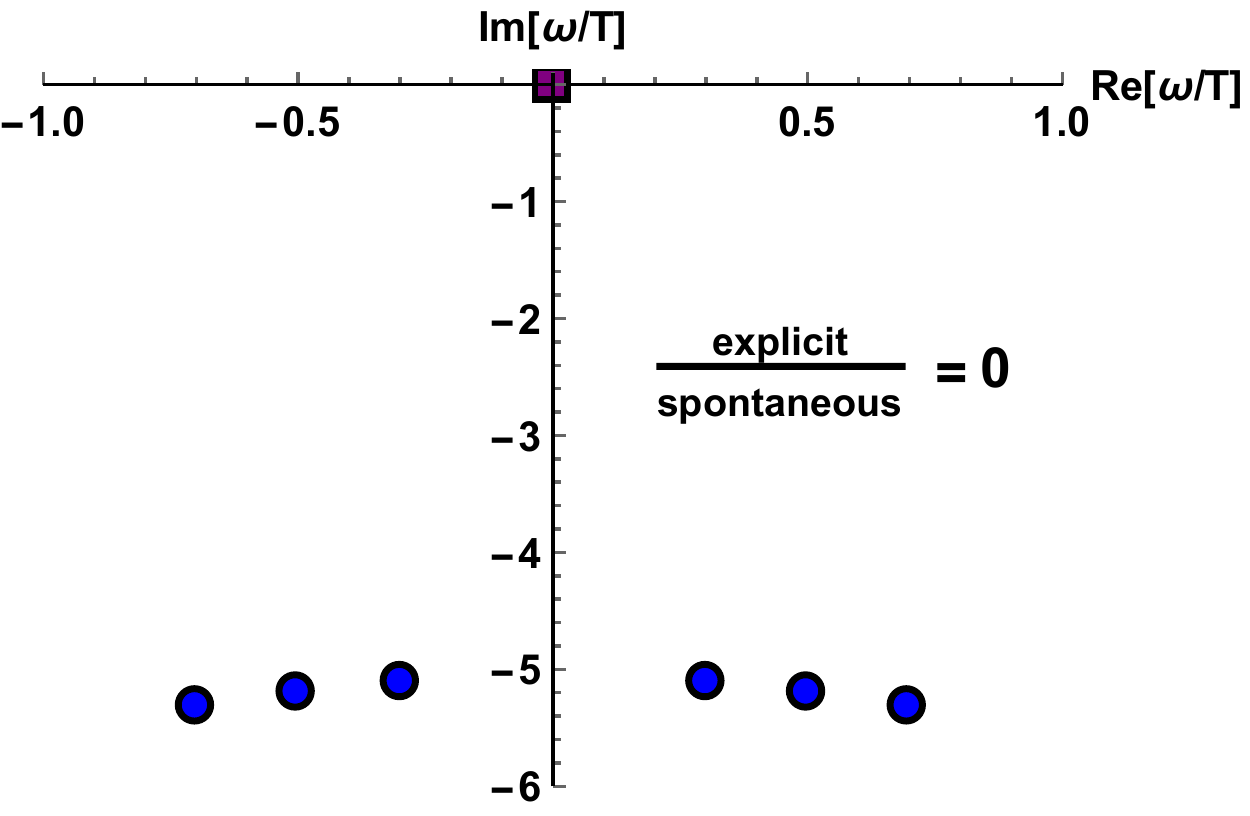}
    \caption{A sketch of Quasi-normal mode spectrum for transverse fluctuations as a function of the ratio between explicit and spontaneous breaking scales. \textbf{Left: }Whenever the spontaneous breaking is negligible, \textit{i.e.} for $\beta \ll 1$, the $\bar{\Omega}$ pole is not an hydrodynamic pole. The only hydrodynamic pole is the Drude pole (if the explicit breaking scale is small, \textit{i.e.} $m/T \ll 1$). \textbf{Center: } In the pseudo-spontaneous case, \textit{i.e.} for $\beta \gg 1$, both the Drude and the $\bar{\Omega}$ poles have small frequency and the expression \eqref{modes} holds. \textbf{Right: }In the purely spontaneous case, \textit{i.e.} for $\beta=\infty$, the Drude and $\bar{\Omega}$ poles collapse into the origin of the complex frequency plane and they produce the double pole identified as the massless phonon.}
    \label{sketch}
\end{figure}

Interestingly, the ratio between the explicit and spontaneous scales corresponds to the ratio between the graviton mass at the UV boundary and the graviton mass at the IR horizon as introduced in \cite{Alberte:2017cch}. More precisely for large $\beta$:
\begin{equation}
    \frac{m_{UV}^2}{m_{IR}^2}\,=\,\frac{m^2}{m^2\,\left(1\,+\,N\,\beta\,u_h^{3+N}\right)}\,\sim\,\frac{1}{\beta}\,=\,\left(\frac{\langle EXB \rangle}{\langle SSB \rangle}\right)^2\,.
\end{equation}
In the scenario where $\beta \gg 1$ the presence of a weakly gapped Goldstone boson is guaranteed (see fig. \ref{sketch}). More specifically, as already suggested in the literature, the presence of such a mode is a consequence of the collision between the Drude pole $\omega\,\sim\,-\,i\,\Gamma$ and a secondary pole $\omega\,\sim\,-\,i\,\bar{\Omega}$ which appears to be light whenever the spontaneous scale is large compared to the explicit one. Concretely, the structure of the low energy excitations at zero momentum can be derived from solving the simple expression \cite{Delacretaz:2017zxd}:
\begin{equation}
    \left(\Gamma\,-\,i\,\omega\right)\,\left(\bar{\Omega}\,-\,i\,\omega\right)\,+\,\omega_0^2\,=\,0 \, ,
\end{equation}
giving rise to the two modes:
\begin{equation}
    \omega_{\pm}\,=\,-\,\frac{i}{2}\,\left(\bar{\Omega}\,+\,\Gamma\right)\,\pm\,\frac{1}{2}\,\sqrt{4\,\omega_0^2\,-\,\left(\Gamma\,-\,\bar{\Omega}\right)^2}\,.\label{modes}
\end{equation}
Let us emphasize that the previous expression \eqref{modes} is obtained using hydrodynamic techniques and it is valid only when the frequencies of the two modes are small compare to the temperature of the system, \textit{i.e.} $\omega_{\pm}/T\,\ll\,1$.
 
In this manuscript we analyze the quasi-normal modes and in particular of their dependence on the explicit, spontaneous and temperature scales. In particular, we investigate the nature of the parameter $\bar{\Omega}$. The nature of this coefficient is still unclear in the literature. The authors of \cite{Andrade:2018gqk} suggested that that parameter is $\bar{\Omega}\,=\,\gamma_1\,\omega_0^2$ where $\gamma_1$ is the coefficient entering in the $\mathcal{G}_{J\phi}$ Green function and $\omega_0$ is the pinning frequency; contrarily the authors of \cite{Amoretti:2018tzw} claimed that $\bar{\Omega}\,=\,M\,\xi$ where $M$ is the mass of the phonon and $\xi$ the Goldstone diffusion constant entering the $\mathcal{G}_{\phi\phi}$ Green function. This existing disagreement represents a further motivation for the analysis pursued in this section.

Let us first discuss the qualitative behaviour of the modes in function of the ratio between explicit breaking and spontaneous breaking, see fig. \ref{sketch}. In the limit of $\langle EXB \rangle \gg\langle SSB \rangle $ and $\langle EXB \rangle\ll 1$, the Drude pole $\Gamma$ is underdamped and it is controlled by the amount of explicit breaking while the $\bar{\Omega}$ pole is overdamped and its imaginary value is much bigger than the Drude pole one. Moreover, it is not singled out but it lives in the sea of the non-hydrodynamic overdamped modes, see fig. \ref{sketch}. On the contrary, in the limit where the breaking becomes pseudo-spontaneous \eqref{pseudodef}, for $\beta \gg 1$, the $\bar{\Omega}$ mode becomes less and less damped and enters in the hydrodynamic regime $\omega/T \ll 1$ along with the Drude pole. This is the limit where the equation \eqref{modes} holds. The two modes at a certain point collide and produce the damped and gapped light phonon which we see in the spectrum. Notice that this collision is also present in the limit of $\langle EXB \rangle \gg\langle SSB \rangle $ but it appears beyond the hydrodynamic limit producing the so-called coherent-incoherent transition \cite{Davison:2014lua}. If we push this further and we assume the  purely spontaneous case, both the Drude and the $\bar{\Omega}$ pole go towards the origin of the complex frequency plane and they form the double pole of the massless phonon observed in \cite{Alberte:2017oqx}. 

\begin{figure}[hbtp]
    \centering
    \includegraphics[width=7.7cm]{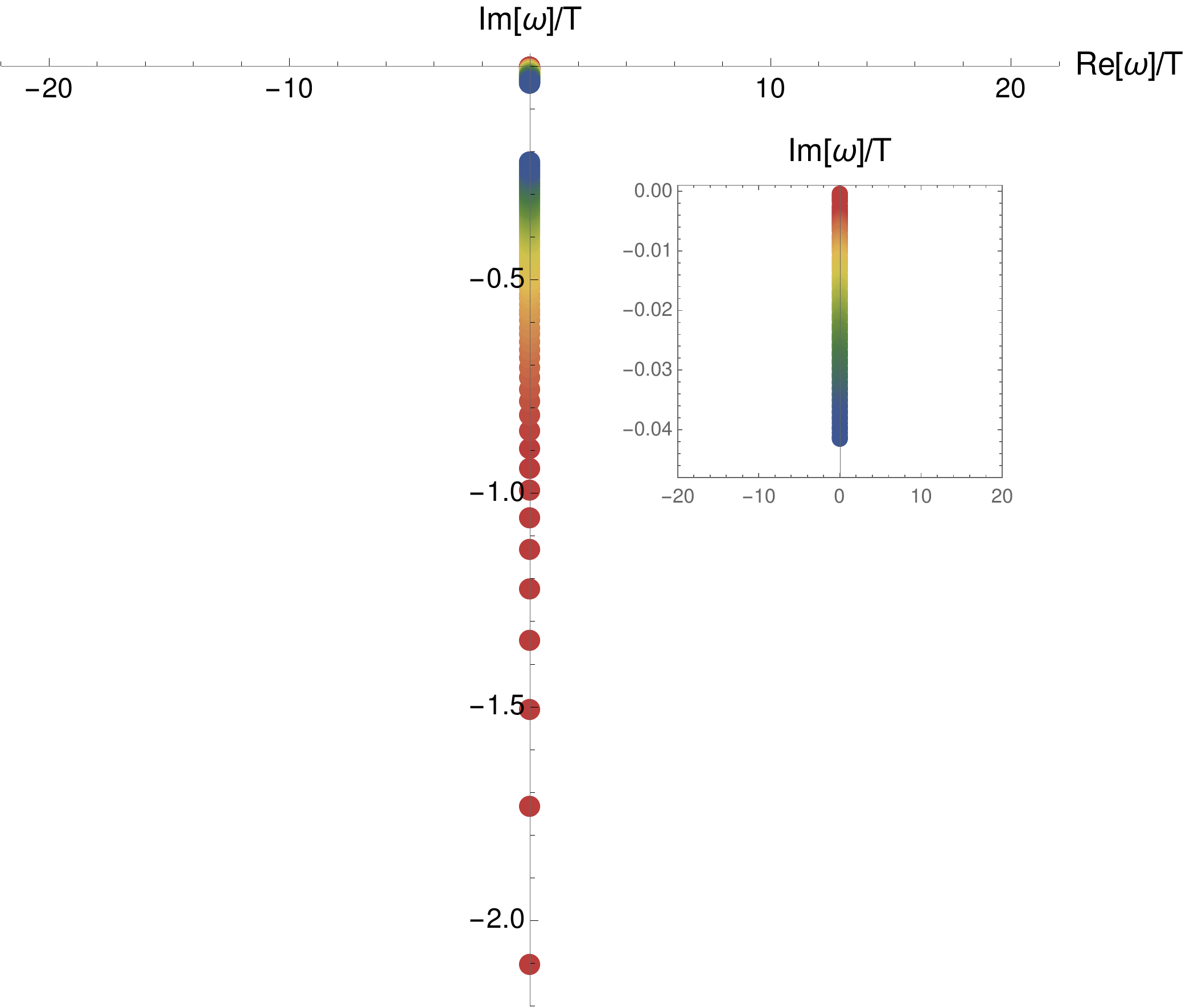}
     \quad
    \includegraphics[width=7.3cm]{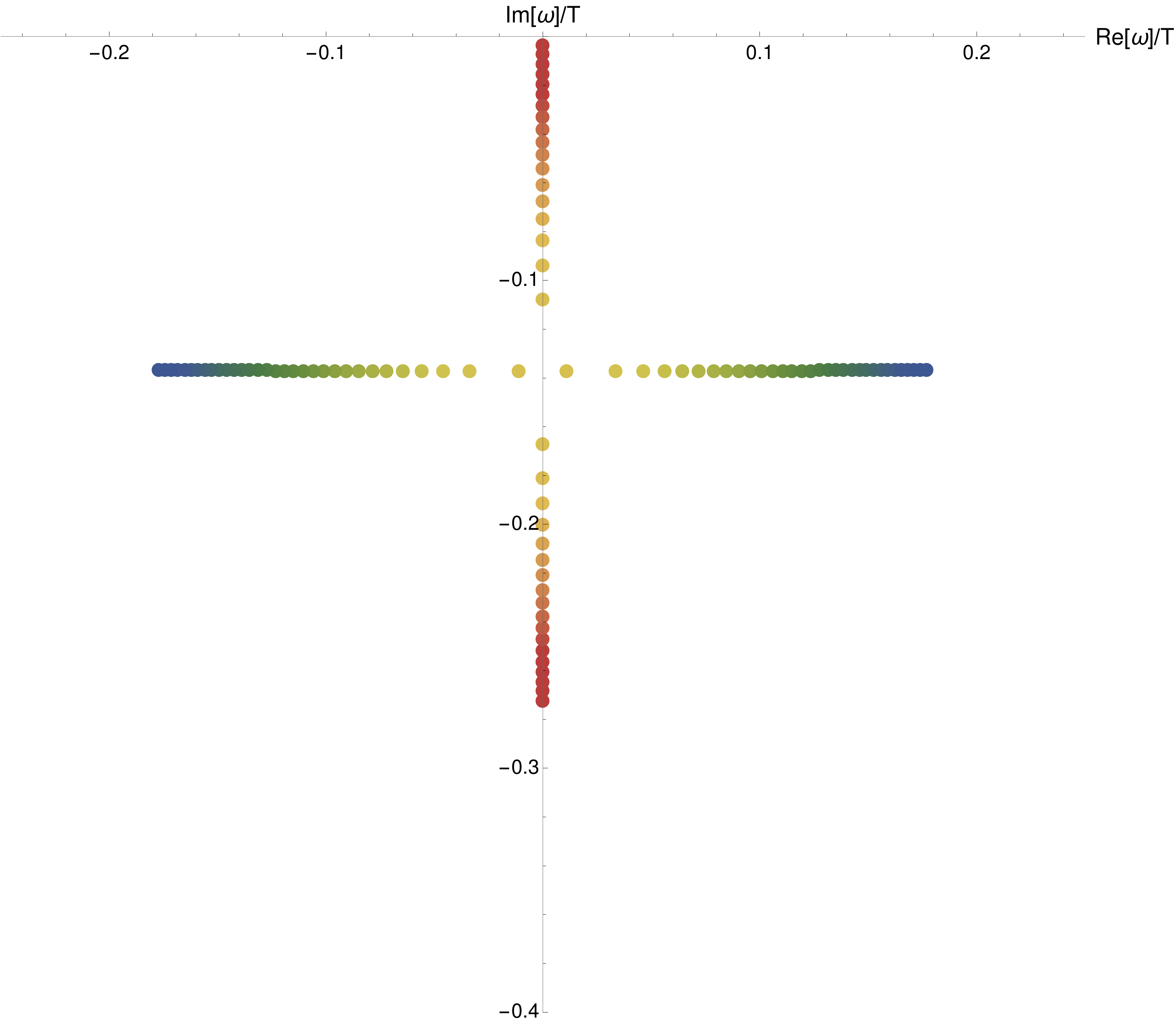}
    \caption{Lowest QNMs at zero momentum for \textbf{Left}: fixed $m/T=0.03$ and $\beta\in[0.5-50]$ (red-blue). \textbf{Right}: Fixed $\beta=50$ and increasing $m/T\in[0.01,0.07]$}.
    \label{fig:first}
\end{figure}

Given the schematic picture regarding the dynamics of the collective modes, now we turn into analyzing the concrete data by determining numerically the QNMs, see \ref{App1} for more details. First, we study how the modes move increasing the ratio between explicit breaking and spontaneous breaking. In order to do that, we fix a small explicit breaking $\langle EXB \rangle /T^2 = m^2/T^2 \ll 1$ and we increase the $\langle EXB \rangle/ \langle SSB \rangle$ moving the dimensionless parameter $\beta$. The results are shown in the left panel of fig. \ref{fig:first}. The lowest pole appears very close to the origin due to the fact that momentum dissipation is negligible in this limit, \textit{i.e.} $\Gamma \sim 0$. Crucially, the second pole, the one governed by the $\bar{\Omega}$ parameter, moves upwards towards the origin increasing the parameter $\beta$. In other words, the more spontaneous the nature of the breaking, the closer to the origin the second pole. At large values of $\langle EXB \rangle/ \langle SSB \rangle$ the second pole is overdamped and far from the hydrodynamic window $\omega/T,k/T \ll 1$. Moving towards the pseudo-spontaneous limit \eqref{pseudodef}, the second pole becomes underdamped and will produce together with the first pole the collision giving rise to the light pseudo-phonon mode.

We can now assume the breaking to be mostly spontaneous, \textit{i.e.} $\beta \gg 1$, and move the values of $m/T$ keeping it rather small. This is done in the right panel of fig. \ref{fig:first}. Notice that in this case, the collision happens at very small values of frequency, close to the origin. Additionally, increasing the values of the explicit breaking, the two poles undergo a collision and they move off-axes. This is the direct manifestation of the fact that the pinning frequency $\omega_0$ grows with the explicit breaking and the pseudo-phonons become more and more massive with it following the GMOR relation \cite{PhysRev.175.2195}.\\

\begin{figure}[hbtp]
    \centering
    \includegraphics[width=7.5cm]{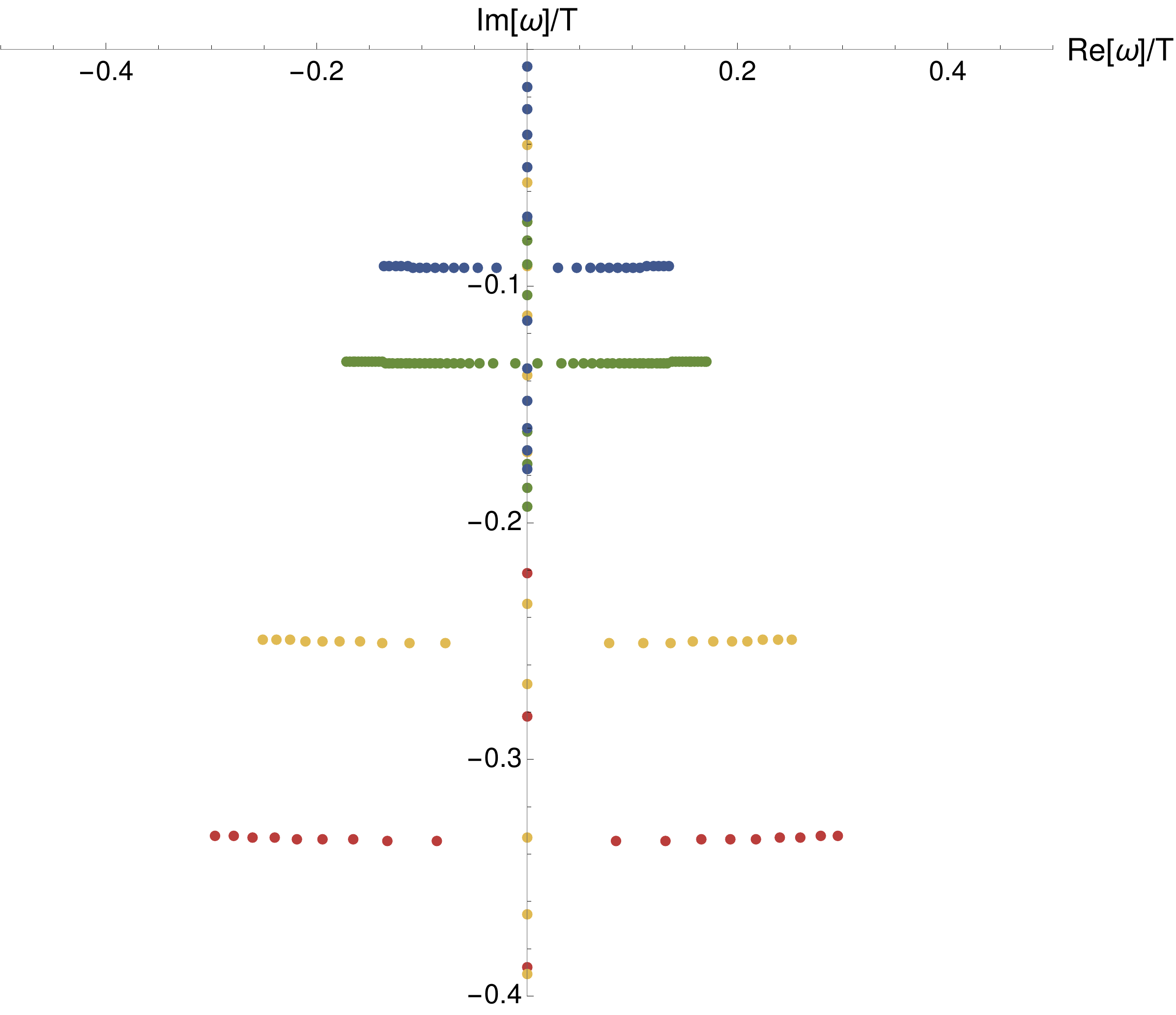}
     \quad
     \includegraphics[width=7.5cm]{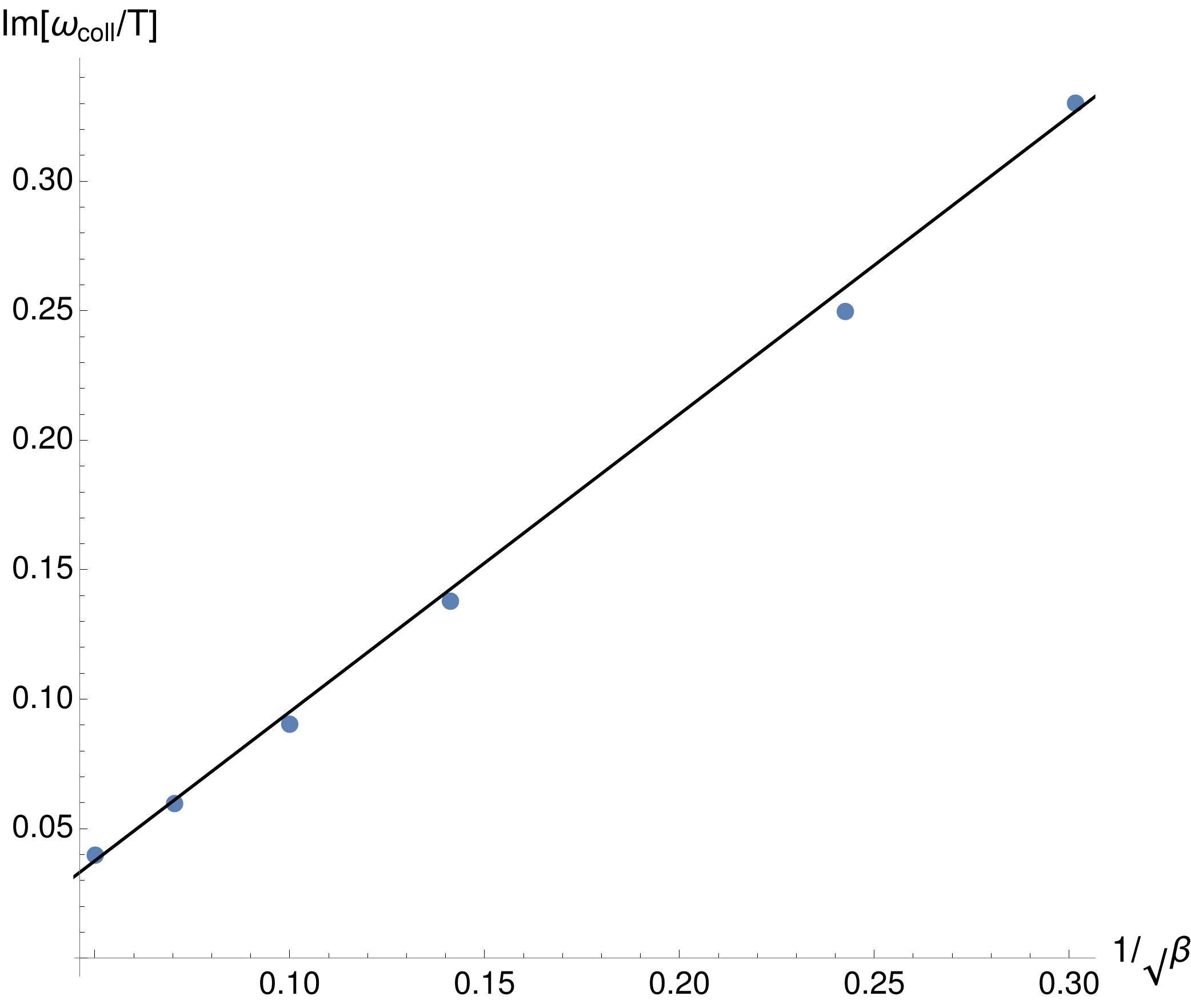}
    \caption{\textbf{Left}:Lowest QNMs for three different values of $\beta=11,17,50,100$ (red-blue) increasing $m/T$ in the vicinity of the collision. \textbf{Right:} Value of the frequency at the collision as a function of $\beta=\langle SSB \rangle/\langle EXB \rangle$.}
    \label{fig:collisions}
\end{figure}

As we already mentioned, the key parameter for the analysis is the ratio between the explicit and the spontaneous breaking which is encoded in the dimensionless scale $\beta$. It is important to notice that the collision between the two poles, controlled respectively by $\Gamma$ and $\bar{\Omega}$, happens for any finite value of $\beta$. The difference is the frequency at which such collision happens. This is shown in fig. \ref{fig:collisions}. When $\beta$ is rather small, and the breaking is mostly explicit, the collision happens at very large values of the frequency. This has already been analyzed in the past as the coherent-incoherent crossover \cite{Davison:2014lua}. Increasing the value of $\beta$, the collision moves upwards towards the origin of the complex plane. Once $\beta$ is very large, the collision happens very close to the origin and it has important consequences on the low energy dynamics of the system. The poles collision produces a very light mode, the pseudo-phonon, which is relevant for the hydrodynamic description. In the extreme limit where the breaking is purely spontaneous, $\beta=\infty$, both the poles lie on the origin and the ``collision'' happens directly from there. As shown in the right panel of figure fig. \ref{fig:collisions}. The collision frequency $\omega_{coll}$ appears to be linearly proportional to the dimensionless parameter $1/\sqrt{\beta}$ which implies the scaling relation
\begin{equation}
  \boxed{ \frac{\omega_{coll}}{T}\,\sim\,\frac{\langle EXB \rangle}{\langle SSB \rangle}\,}
\end{equation}
\section{A new phase relaxation mechanism}
Let us study the behaviour of the parameters $\omega_0,\bar{\Omega}$ as a function of the $\langle EXB \rangle$ and $\langle SSB \rangle$ scales by tuning $m/T$ and $\beta$ accordingly.

\begin{figure}[hbtp]
    \centering
    \includegraphics[width=7.3cm]{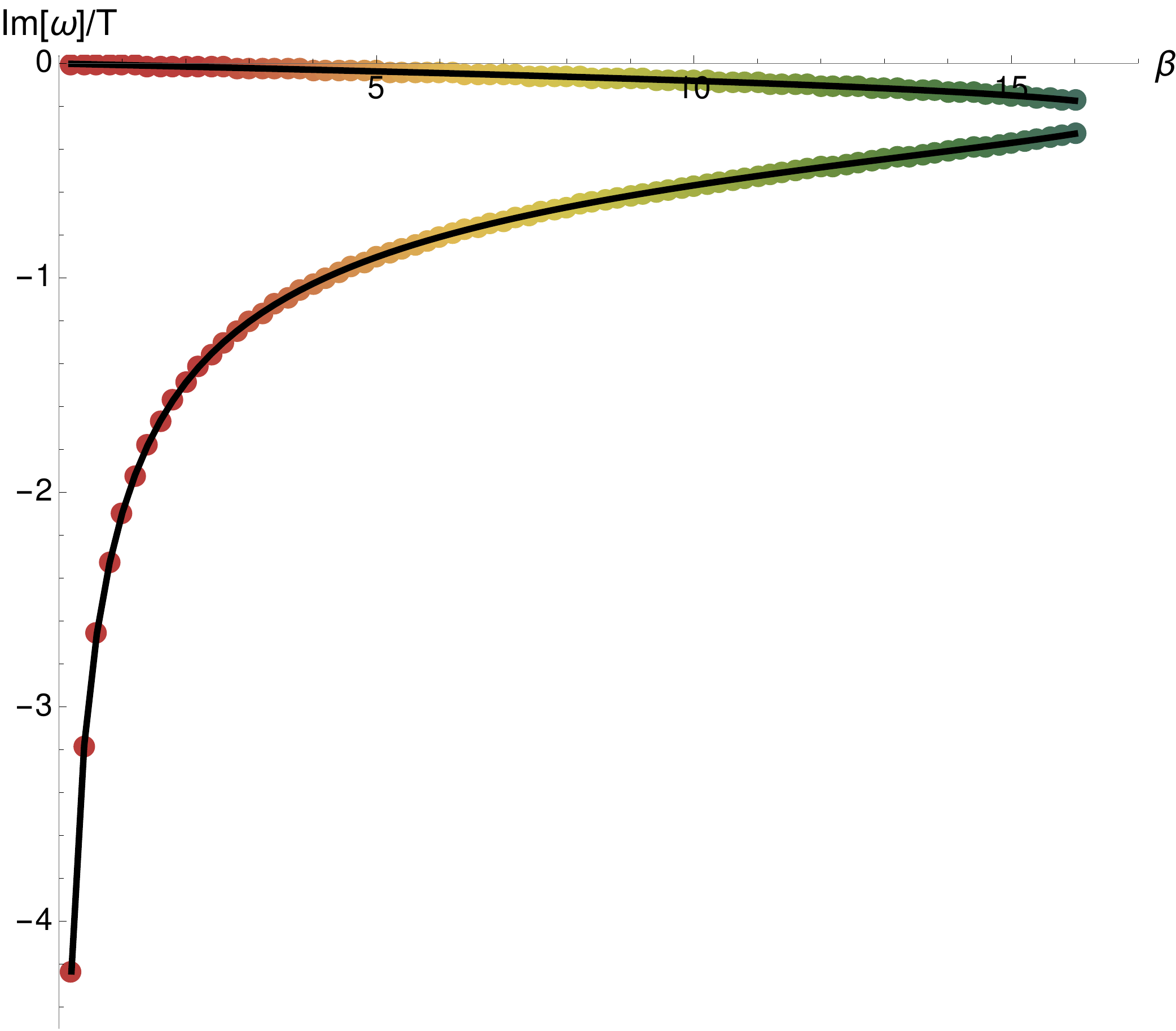}
     \quad
     \includegraphics[width=7.7cm]{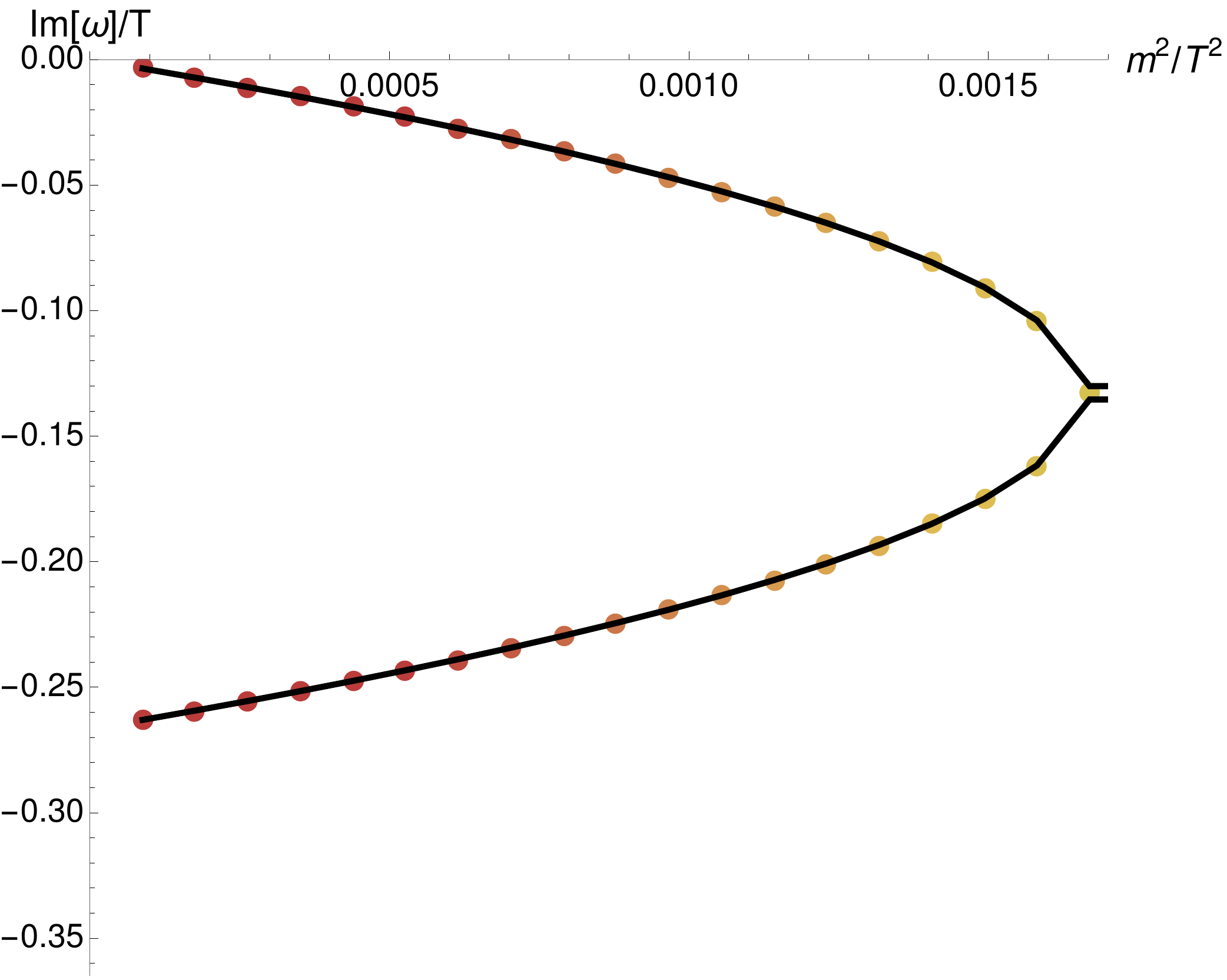}
    \caption{Fits (black lines) compared to data (colored dots) corresponding to the fits, shown in figures \ref{fig:second} and \ref{fig:third}, for data plotted in figure \ref{fig:first}. }
    \label{fig:fit}
\end{figure}

\begin{figure}[hbtp]
    \centering
  
   \includegraphics[width=0.48 \linewidth]{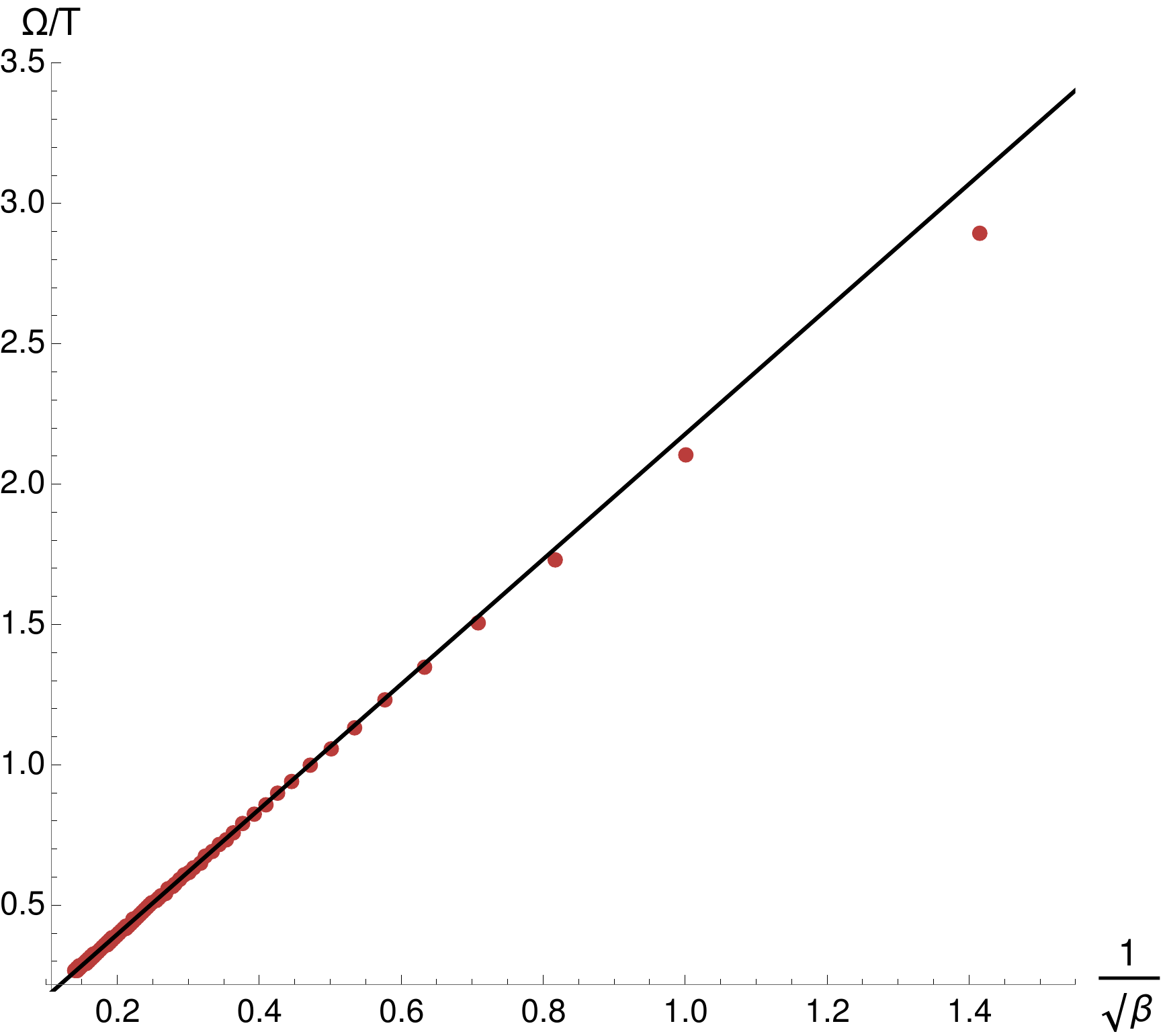}%
    \quad
 \includegraphics[width=0.48 \linewidth]{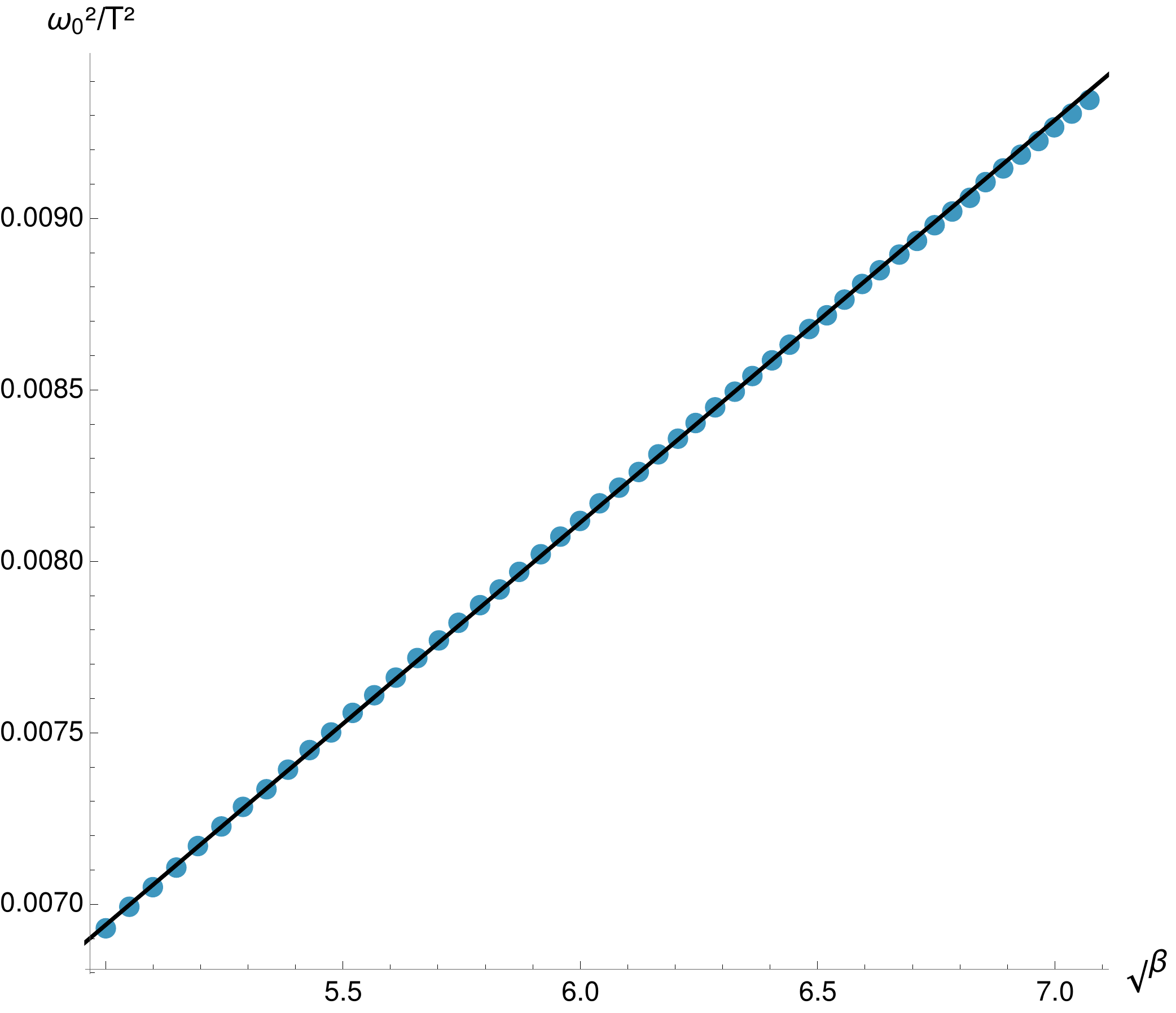}
    \caption{\textbf{Left:} Fitted $\Omega/T$ for fixed $m/T=0.01$. \textbf{Right:} Fitted $\omega_0^2/T^2$ for fixed $m/T=0.03$. In both cases the deviation from the expected behaviour appears at small $\beta$, i.e. away from the pseudo-spontaneous limit.}
    \label{fig:second}
\end{figure}
Combining hydrodynamic and holographic arguments (see \cite{Davison:2013jba,Amoretti:2018tzw} for more details) we can derive a simple analytical formula:
\begin{equation}
    \Gamma\,+\,\frac{\omega_0^2}{\bar{\Omega}}\,=\,m^2\,\frac{V_X}{2\,\pi\,T}\,+\,\mathcal{O}(m^4)\, .\label{dd}
\end{equation}
For our potential $V(X)=X+\beta X^N$, from the results in \cite{Alberte:2017oqx} and \cite{Davison:2013jba} the most natural splitting is:
\begin{equation}
    \Gamma\,=\,\frac{m^2}{2\,\pi\,T}\,\sim\,\langle EXB \rangle ^2\,+\,\dots,\quad \quad \frac{\omega_0^2}{\bar{\Omega}}\,=\,\frac{m^2\,\beta\,N}{2\,\pi\,T}\,\sim\,\langle SSB \rangle^2\,+\,\dots \label{new}
\end{equation}

As we will see later, these scalings are in agreement with our numerical data.\\
In order to fit them we use formula \eqref{modes} supplemented by the definition of the momentum dissipation rate at small explicit breaking $m/T \ll 1$. Notice that for $m/T \ll 1$ we immediately have $\Gamma/T \ll 1$, which means the momentum dissipation rate can be neglected. That said, we proceed to fit  the numerical data. The quality of the fits is graphically shown in fig. \ref{fig:fit}. We study the parameters $\bar{\Omega},\omega_0$ in function of the explicit and spontaneous scales. First we plot the dependence on the dimensionless parameter $\beta$ in fig. \ref{fig:second}. The find the pinning frequency $\omega_0^2$ to be proportional to $\sqrt{\beta}$. More interestingly, the new parameter $\bar{\Omega}$ is proportional to the inverse of $\sqrt{\beta}$, the ratio between the explicit breaking scale \eqref{expscale} and the spontaneous one \eqref{spontscale}.

\begin{figure}
    \centering
      \includegraphics[width=7.5cm]{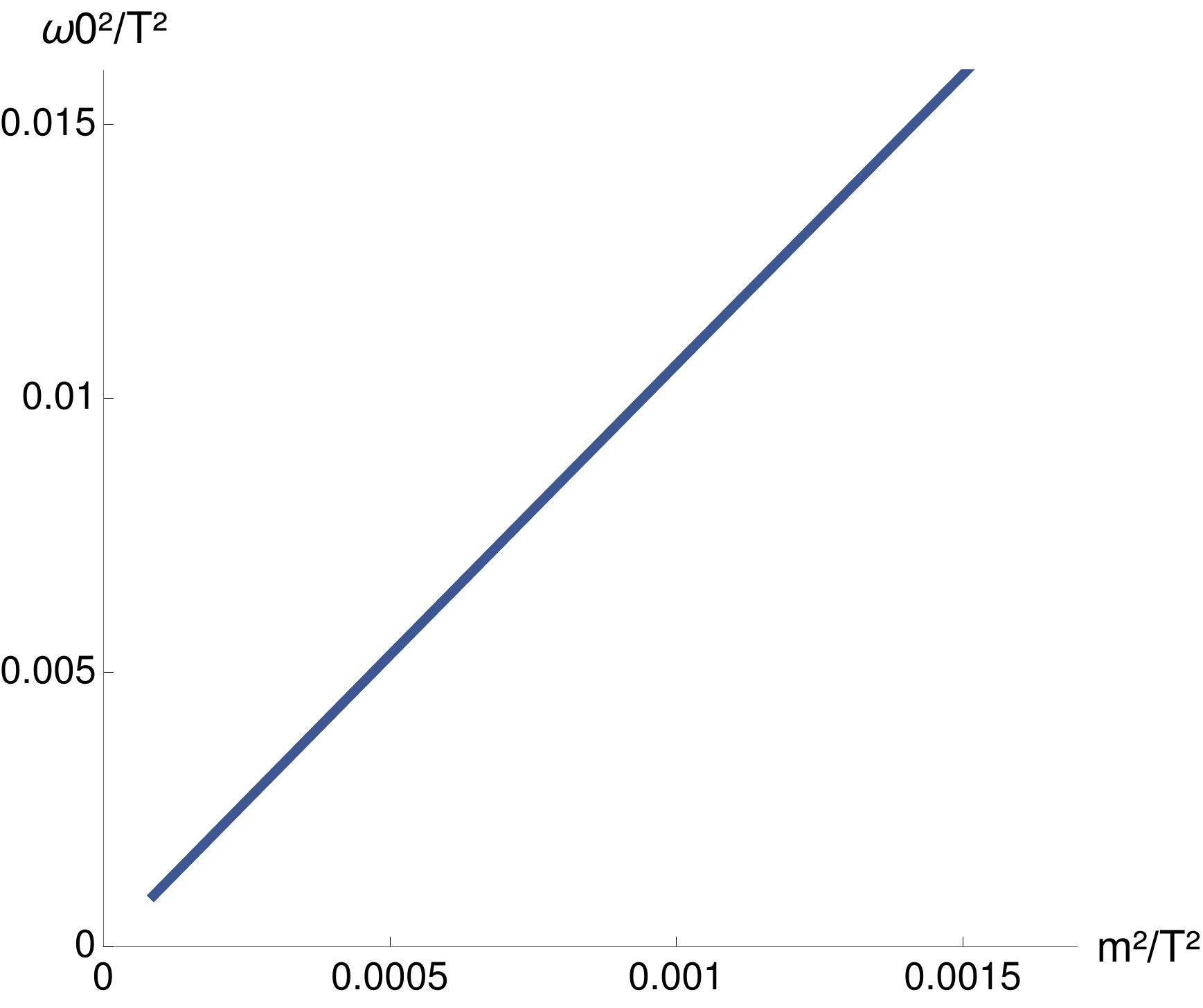}
    \quad
   \includegraphics[width=7.5cm]{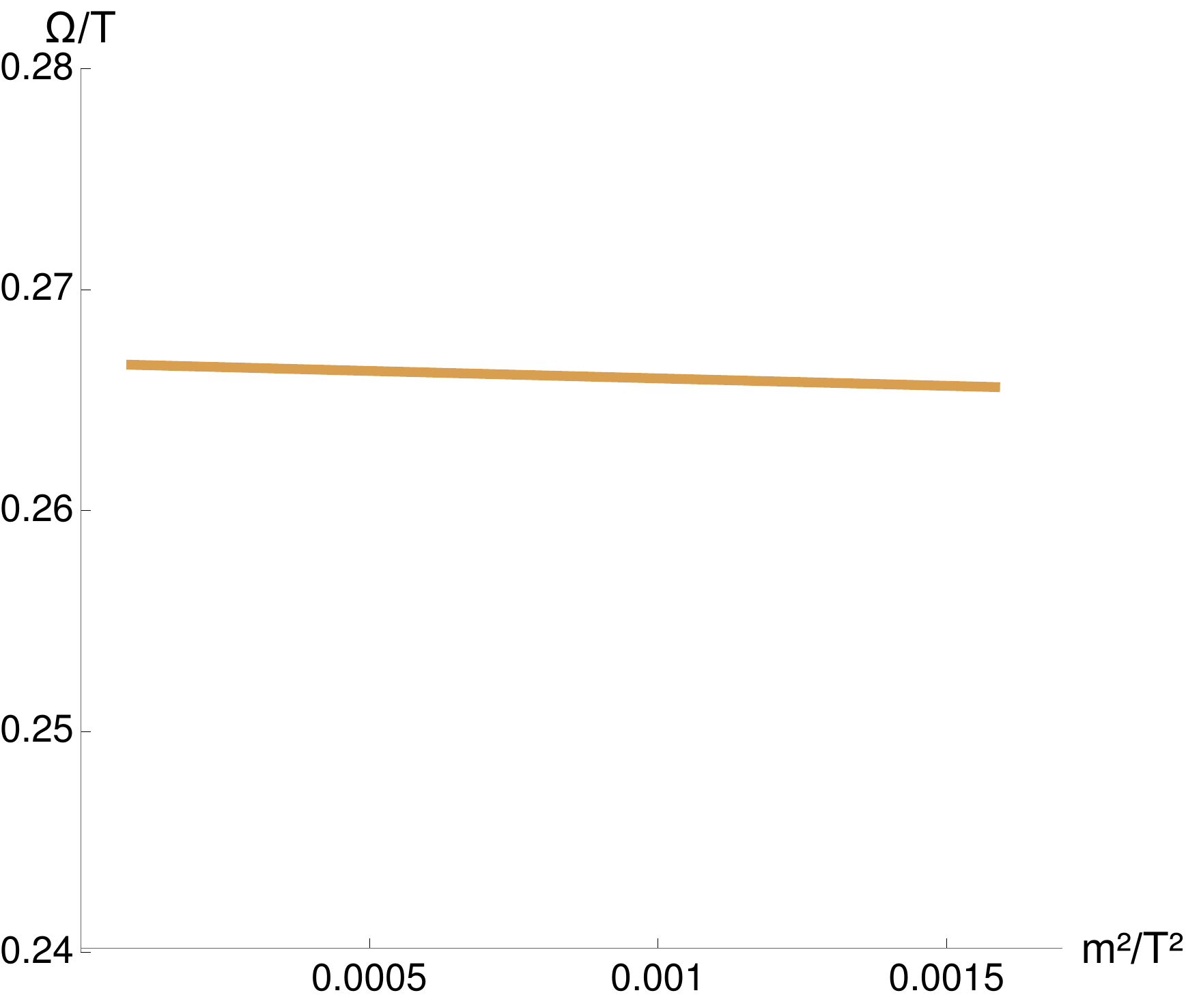}
       \caption{Dependence of the fitted $\Omega/T$ and $\omega_0^2/T^2$ on $m^2/T^2$ for fixed $\beta=50$ .}
    \label{fig:third}
\end{figure}

We also show the dependence of the parameters in function of $m$ in fig. \ref{fig:third}. Here the pinning frequency $\omega_0^2$ is proportional to the mass square $m^2/T^2$. The dependence of $\omega_0$ on $\beta$ and $m/T$ proves the validity of the GMOR relation in our model. 
Furthermore we see that at leading order in $m^2/T^2$
\begin{equation}
    \bar{\Omega}\,\sim\,\frac{\langle EXB \rangle }{\langle SSB \rangle }\,\left(1\,+\,\mathcal{O}\left(\langle EXB \rangle \right)\,\right)\, ,
\end{equation}
where the higher order terms are suppressed by powers of $\langle EXB \rangle$, which is small in the pseudo-spontaneous limit \eqref{pseudodef}.

The previous analysis allows us to come to the main point of this section. In the pseudo-spontaneous regime \eqref{pseudodef}, we find:
\begin{equation}
    \boxed{ \omega_0^2\sim\,\langle EXB \rangle\langle SSB\rangle,\quad \bar{\Omega}\sim \frac{\langle EXB \rangle}{\langle SSB \rangle}\, ,} \label{main}
\end{equation}
which is our main result. This provides a complete picture of the pseudo-spontaneous regime and the parameters appearing in the effective hydrodynamic description as a function of the explicit and spontaneous breaking scales. \\

Let us summarize what we found:
\begin{enumerate}
    \item The relation for the pinning frequency $\omega_0$ is the well-known GMOR relation for a pseudo-Goldstone boson \cite{PhysRev.175.2195}.
    \item The relation between the relaxation parameter $\bar{\Omega}$ and the explicit/spontaneous breaking scales, eq.\ref{main}, is new and important. First, from this result, we notice that the effect of $\bar{\Omega}$ is not only related on the spontaneous breaking scale of translations. It depends crucially on the explicit scale and is incompatible with the interpretation of it as a phase relaxation mechanism \textit{stricto sensu}. Moreover, we see that in the case of purely spontaneous breaking the parameter $\bar{\Omega}$ vanishes and hence gives rise to the appearance of the transverse sound mode. On the contrary, in the mostly explicit case this mode is overdamped and does not participate in the low energy dynamics.
    \item The frequency at which the two lowest modes collide is indeed governed by the new phase relaxation scale:
    \begin{equation}
        \frac{\omega_{coll}}{T}\,\sim\,\frac{\bar{\Omega}}{T}
    \end{equation}
\end{enumerate}
We conclude this section by considering the relation proposed in \cite{Amoretti:2018tzw} which relates this novel phase relaxation scale, the mass of the pseudo-goldstone modes $\mathrm{M}$ and the goldstone diffusion constant $\xi$. More precisely, it is conjectured that in the pseudo-spontaneous limit:
\begin{equation}
 \label{uni}   \bar{\Omega}\,\sim\,\mathrm{M}^2\,\xi\,=\,\frac{\omega_0^2\,\chi_{PP}}{G}\,\xi
\end{equation}
Let us emphasize that the Goldstone diffusion constant can be obtained numerically from the $\mathcal{G}_{\Phi\Phi}$ correlator. More details can be found in appendix \ref{app2}. Moreover, it is possible to derive an horizon formula \cite{Amoretti:2019cef} for this transport coefficient which in our case reads:
\begin{equation}
    \frac{\xi}{G}\,=\,\frac{4\,\pi\,s\,T^2}{2\,m^2\,\chi_{PP}^2\,V_X}\label{eq:more}
\end{equation}
and which has been confirmed with the numerical results for our model in \cite{Ammon:2019apj}.

\begin{figure}[h!]
    \centering
     \includegraphics[width=7.5cm]{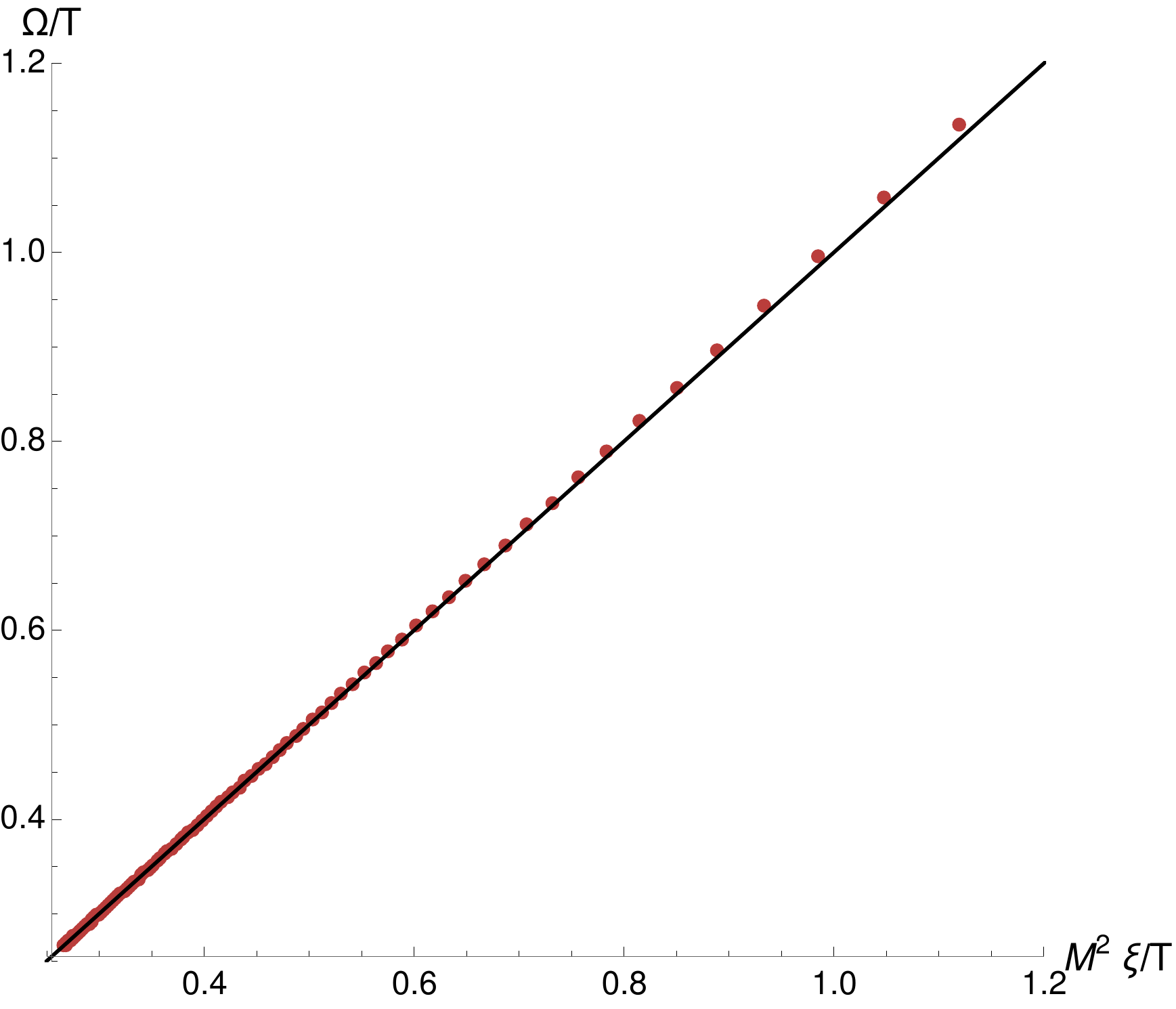}
    \quad
         \caption{The numerical check of the relation in eq.\eqref{uni}.}
    \label{fig:scaling}
\end{figure}

Let us first explain why our result $\bar{\Omega}\,\sim \langle EXB \rangle / \langle SSB \rangle$ already implies the validity of the relation \eqref{uni}. Just following the scalings we derived in the previous sections we immediately obtain\footnote{Keep in mind that in terms of the breaking scales $\chi_{PP}\sim \mathcal{O}(0)$.}:
\begin{equation}
  \underbrace{\frac{\langle EXB \rangle}{\langle SSB \rangle}}_{\bar{\Omega}}\,\sim\,\underbrace{\langle EXB \rangle \langle SSB \rangle}_{\omega_0^2}\,\underbrace{\frac{1}{\langle SSB \rangle^2}}_{\xi/G}\,\quad
\end{equation}
which just prove a relation like eq.\eqref{uni} should hold.\\
We can do a second step and use the analytic formula for the ratio $\omega_0^2/\bar{\Omega}$ defined in eq.\eqref{new}. By putting all our results together, we obtain analytically that in the limit $m/T \ll 1$ we have:
\begin{equation}
    \frac{\bar{\Omega}\,G}{\omega_0^2\,\chi_{PP}\,\xi}= \frac{1\,+\,N\,\beta}{N\,\beta}\,\underbrace{\longrightarrow}_{\beta \rightarrow \infty}\,1
\end{equation}
which holds in the pseudo-spontaneous limit. Here we used that in the limit $m/T \ll 1$:
\begin{equation}
    \chi_{PP}\,=\,\epsilon\,+\,p\,=\,s\,T\,+\,\mathcal{O}(m^2)\, .
\end{equation}

The analytic result is confirmed by the numerical data. See for example fig.\ref{fig:scaling}.\\
In conclusion we have been able to prove analytically that in the pseudo-spontaneous limit:
\begin{equation}
    \boxed{\bar{\Omega}\,=\,\mathrm{M}^2\,\xi\,=\,\frac{\omega_0^2\,\chi_{PP}}{G}\,\xi}
\end{equation}
which was conjectured in \cite{Amoretti:2018tzw}.
\section{Conclusions}\label{sec: concl}
In this work we provide a complete and unified picture of the breaking of translational invariance in simple holographic models. By ``simple'' we mean holographic models where translations are broken but the background geometry is homogeneous. This is possible thanks to the preservation of a diagonal symmetry group built from the spacetime translations and the internal symmetries of the Stueckelberg fields \cite{Alberte:2015isw}.\\
The first aim of this paper is to resolve the confusion in the literature regarding the different explicit and spontaneous breaking mechanisms and their interplay. Once compared the holographic results with the effective field theory and hydrodynamic predictions, we discuss the nature of the $\bar{\Omega}$ relaxation scale which appears in the pseudo-spontaneous picture. Such a parameter appeared long time ago in \cite{Alberte:2017cch}, but it has been recently topic of discussion in the literature \cite{Amoretti:2018tzw,Andrade:2018gqk}. More precisely, it has been claimed in \cite{Amoretti:2018tzw} that its nature is related to phase relaxation given by topological defects and the presence of fluctuating order. In our work we suggest that is not the case \textit{stricto sensu}. We present the following arguments:
\begin{enumerate}
    \item The frequency dependent viscosity in the SSB limit does not display any Drude peak.
    \item The crystal diffusion mode in the SSB limit is not damped. This has been shown by some of the authors of this work in two preprints \cite{Ammon:2019apj,Baggioli:2019abx} appeared at the same time of our work.
    \item The ``novel'' parameter $\bar{\Omega}$ depends linearly on the explicit breaking scale.
\end{enumerate}
Nevertheless, it is definitely true, as observed in \cite{Amoretti:2018tzw,Andrade:2018gqk}, that this parameter enters in the definition of the transverse collective modes and the electric conductivity in the same way as a proper phase relaxation would. However, it is not clear yet if it does play an analogous role in the shear correlator. Even in the pseudo-spontaneous limit we do not observe any Drude pole in the frequency dependent viscosity $\eta(\omega)$. This result does not a priori contradicts the results of \cite{Delacretaz:2017zxd} but it emphasizes the need of constructing a complete generalization in presence of explicit breaking. The latter should encode a new mechanism dependent on the explicit breaking scale\footnote{A different phase relaxation mechanism produced by disorder has been already discussed in the literature \cite{doi:10.1143/JPSJ.45.1474,fogler2000dynamical}. That could be compatible with what we observe here in the parameter $\bar{\Omega}$. We are grateful to Blaise Gouteraux for mentioning this point to us.}.\\
To make progress in this direction we prove in this manuscript that:
\begin{equation}
    \bar{\Omega}\,\sim\,\frac{\langle EXB\rangle}{\langle SSB \rangle}\, ,
\end{equation}
and we argue that this holds in general. To the best of our knowledge, this represents a new result which definitely deserves more investigation.\\
Furthermore, we show that this result already is equivalent to the relation conjectured in \cite{Amoretti:2018tzw} between the phase relaxation scale, the mass of the pseudo-Goldstone modes $\mathrm{M}$ and the Goldstone diffusivity $\xi$. More precisely, we have been able to prove analytically that within our class of holographic models $V(X)=X+X^N$:
\begin{equation}
    \bar{\Omega}\,=\,\mathrm{M}^2\,\xi\,=\,\frac{\omega_0^2\,\chi_{PP}\,\xi}{G}
\end{equation}
and to show that this formula is in perfect agreement with our numerical data\footnote{The constant of proportionality $\bar{\Omega}\,=\,c\,\mathrm{M}^2\,\xi\,=\,c\,\frac{\omega_0^2\,\chi_{PP}\,\xi}{G}$, which in our case is $c=1$, might depend on the details of the model and it is not physically relevant.}. We believe this relation might help in the construction of a complete hydrodynamic theory for the novel phase relaxation scale $\bar{\Omega}$. It represents a further step in the understanding of this new mechanism. We hope to report more about this point in the close future.
\section*{Acknowledgements}
We thank Andrea Amoretti, Tomas Andrade, Blaise Gouteraux, Saso Grozdanov, Karl Landsteiner, Daniele Musso, Napat Poovuttikul, Alessio Zaccone and Weijia Li for several discussions and helfpul comments. We thank Sean Gray and Sebastian Grieninger for collaboration on related projects. We did greatly benefit from several fruitful conversations with our colleague Daniel Arean. M.A. is funded by the Deutsche Forschungsgemeinschaft (DFG, German Research  Foundation) -- 406235073. M.B. acknowledges the support of the Spanish MINECO’s “Centro de Excelencia Severo Ochoa” Programme under grant SEV-2012-0249. A.J. is supported by the ``Atracci\'on del Talento'' program (Comunidad de Madrid) under grant 2017-T1/TIC-5258. M.B. would like to thank A. Jimenez and M. Siouti for teaching him how to be classy and how to control his obnoxious aggressivity.
\appendix
\section{EOMs and Green Functions}\label{App1}
In this appendix we present few relevant technical details regarding the computations discussed in the main text.\\[0.2cm]
\textbf{Equations of motion for the fluctuations}\\[0.1cm]
We consider the following set of perturbations:
\begin{equation}
a_x,\,h_{tx}\equiv u^2 \delta g_{tx},\,h_{xy}\equiv u^2\delta g_{xy},\,\delta \phi_x,\,\delta g_{xu}\, ,
\end{equation}
and we assume for simplicity the radial gauge, \textit{i.e.} $\delta g_{xu}=0$.\\
The corresponding equations of motion are:
\begin{align}
&0=-2(1-u^2\,V''/V')h_{tx}+u\,h_{tx}'-i\,k\,u\, h_{xy}-\left( k^2\,u+2\,i\,\omega (1-u^2\,V''/V')\right)\,\delta \phi_x+u\,f\, \delta \phi_x''
\nonumber\\&+\left(-2(1-u^2\,V''/V')\,f+u\,(2 i \omega+f')\right)\delta \phi_x'
\,\,;\nonumber\\[0.1cm]
&0=2\,i\,m^2\,u^{2}\omega V'\,\delta \phi_x+u^2\, k\,\omega\, h_{xy}+2\,u^4\,\mu\,(i\omega\,a_x+f\,a_x')
\nonumber\\&+\left(6+k^2\,u^2-4\,u^2\mu^2-2 \, m^2 (V-u^2\,V' ) \, -6 f+2uf'\right) \, h_{tx}+\left(2\,u\,f-i\,u^2\omega\right)h_{tx}'-u^2\,f\,h_{tx}'' \,\,;
\nonumber\\[0.1cm]
&0=2i\,k\,u\,h_{tx}-iku^2h_{tx}'-2\,i \, k \,m^2 \,u^{2}V'\delta \phi_x+2 h_{xy}\left(3+i\,u \, \omega-3f+uf'- \,m^2(V-u^2 V') \right)
\nonumber\\&-\left(2i\,u^2 \, \omega-2uf+u^2\,f'\right)h_{xy}'-u^2\,f\, h_{xy}''\,\,;
\nonumber\\[0.1cm]
&0=2\,h_{tx}'-u\,h_{tx}''-2m^2\,u\,V' \, \delta\phi_x'+ik\,u\,h_{xy}'+2u^4 \,\mu \, a_x'\,\,;\nonumber\\[0.1cm]
&0=-\mu\,h_{tx}'-k^2\,a_x+(2i\,\omega+f')\,a_x'+f\,a_x'' \,.\label{eq:equationx}
\end{align}
These are the equations we solve numerically to obtain the correlators and the QNMs discussed in the main text. Notice that since we are using Eddington-Filkenstein coordinates at the horizon we have to impose simply regularity.\\[0.2cm]
\textbf{Asymptotics and Green functions}\\
The UV asymptotics for the perturbations considered are:
\begin{align}
&\delta \phi_x\,=\,\phi_{x\,(l)}\,(1\,+\,\dots)+\,\phi_{x\,(s)}\,u^{5-2n}\,(1\,+\,\dots)\,,\quad\nonumber\\ &h_{tx}\,=h_{tx\,(l)}\,(1\,+\,\dots)\,+\,h_{tx\,(s)}\,u^{3}\,(1\,+\,\dots)\,,\nonumber\\ &h_{xy}\,=h_{xy\,(l)}\,(1\,+\,\dots)\,+\,h_{xy\,(s)}\,u^{3}\,(1\,+\,\dots)\,,\nonumber\\\quad &a_x\,=a_{x\,(l)}\,(1\,+\,\dots)\,+\,a_{x\,(s)}\,u\,(1\,+\,\dots)\,\,\,,
\end{align}
where the boundary is taken to be at $u=0$.\\
Using the standard AdS-CFT dictionary we can then write down the Green functions as:
\begin{align}\label{greenF}
&\mathcal{G}^{\textrm{(R)}}_{T_{tx}T_{tx}}\,=\,\frac{2\,\Delta-d}{2}\,\frac{h_{tx\,(s)}}{h_{tx\,(l)}}\,=\,\frac{3}{2}\frac{h_{tx\,(s)}}{h_{tx\,(l)}}\,,\nonumber\\
&\mathcal{G}^{\textrm{(R)}}_{T_{xy}T_{xy}}\,=\,\frac{2\,\Delta-d}{2}\,\frac{h_{xy\,(s)}}{h_{xy\,(l)}}\,=\,\frac{3}{2}\frac{h_{xy\,(s)}}{h_{xy\,(l)}}\,,\nonumber\\
&\mathcal{G}^{\textrm{(R)}}_{JJ}\,=\,\frac{a_{x\,(s)}}{a_{x\,(l)}}\,,
\end{align}
where the time and space dependences are omitted for simplicity.\\
For the spontaneous case $V(X)=X^N$: 
\begin{equation}
    \mathcal{G}^{\textrm{(R)}}_{\phi\phi}\,=\,N\,m^2\,\left(2N-5\right)\frac{\phi_{x\,(s)}}{\phi_{x\,(l)}}\, .
\end{equation}
All the physical observables analyzed in this letter can be extracted from the retarded Green's functions.\\[0.2cm]
\textbf{Numerical techniques}\\
 We have used pseudospectal techniques to determine the quasi-normal modes of the system \cite{Kokkotas:1999bd,Berti:2009kk}. 
 As in previous works, we have used directly the fluctuations of the non-gauge invariant fields in infalling Eddington-Finkelstein coordinates as shown in \ref{eq:equationx}. The fields are decomposed in Chebyshev polynomials, which ensure regularity at the horizon and automatically force the leading, divergent, terms of the asymptotic expansion to vanish. As discussed in \cite{Ammon:2017ded, Ammon:2016fru}, the problem of finding the quasi-normal modes can be recasted in terms of a generalized eigenvalue problem. This approach has the advantage of directly finding many QNMs with no need of shooting, we refer the reader to \cite{Ammon:2017ded, Ammon:2016fru} for a complete discussion on this subject. All our data have been computed with 50 gridpoints and 60 digits precision. We have checked both convergence and that the equations of motion are satisfied.

\section{The Goldstone correlator}\label{app2}
In this section we consider the Green functions which contain the Goldstone operator $\phi$ dual to the bulk Stuckelberg fields and which are relevant for our discussion. We indicate with $P$ the momentum operator of the dual QFT which is dual to the metric function $h_{ti}$, and the charged current $J$ dual to the bulk gauge field $A_{\mu}$. The correlators can be extracted via standard AdS-CFT techniques using the holographic dictionary \cite{Skenderis:2002wp,Ammon:2015wua,zaanen2015holographic,Hartnoll:2016apf} (see Appendix \ref{App1} for more details).

\begin{figure}[hbtp]
    \centering
    \includegraphics[width=7.5cm]{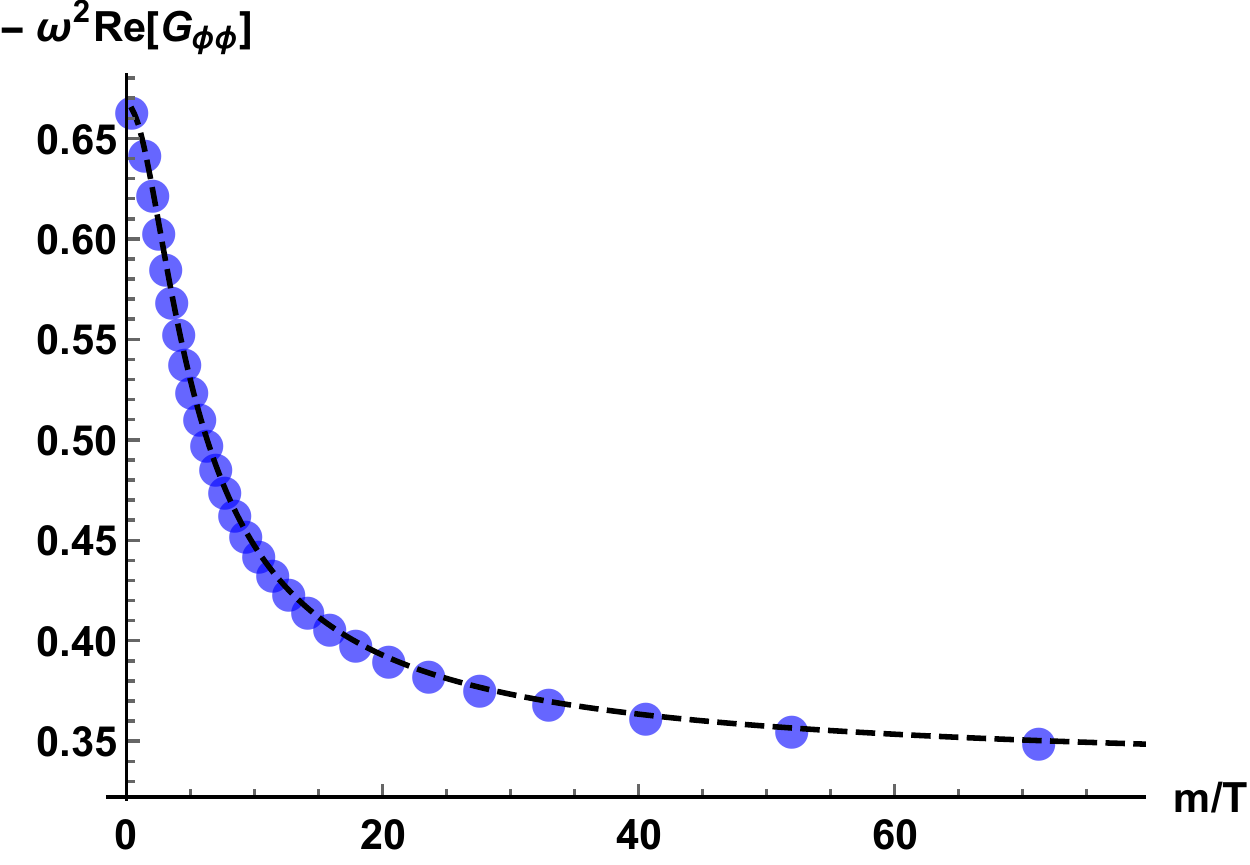}
    \quad
     \includegraphics[width=7.5cm]{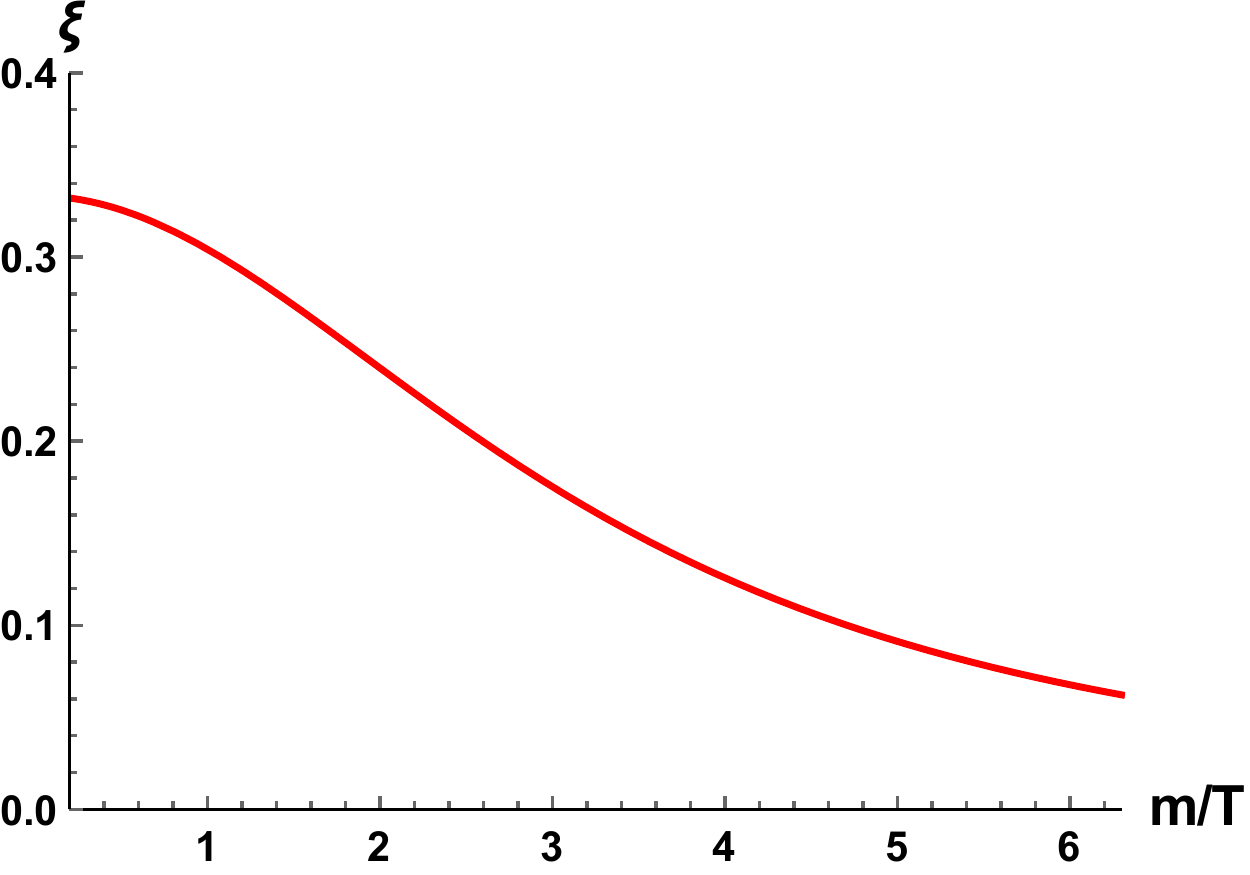}
    \caption{Data for the potential $V(X)=X^3$ and zero chemical potential. \textbf{Left: }Check of the correlator structure \eqref{eqdue}. The blue dots are numerical data and the dashed line is the analytic result for $\chi_{PP}^{-1}$. \textbf{Right: } The $\xi$ parameter extracted from the imaginary part of the $\phi$ correlator.}
    \label{figG}
\end{figure}
In the spontaneous breaking limit\footnote{In the pseudo-spontaneous case there will appear corrections from the explicit breaking strength, proportional to the momentum relaxation rate $\Gamma$. Nevertheless, as far as the explicit breaking scale is not large, such corrections will be negligible and the expressions \eqref{equno},\eqref{eqdue},\eqref{e1tre} will still be a good approximation for the correlators.}, the low energy (small frequency) expansions can be deduced using hydrodynamic methods \cite{PhysRevB.22.2514, Delacretaz:2017zxd} and take the simple forms:
\begin{equation}
    \mathcal{G}_{J\phi}(\omega)\,\equiv\,\langle\,J\,\phi\,\rangle\,=\,\gamma_1\,+\,\frac{\rho}{\chi_{PP}}\,\frac{i}{\omega}\, ,\label{equno}
\end{equation}
\begin{equation}
    \mathcal{G}_{\phi\phi}(\omega)\,\equiv\,\langle\,\phi\,\phi\,\rangle\,\,=\,-\,\frac{1}{\omega^2\,\chi_{PP}}\,+\,\frac{\xi}{G}\,\frac{i}{\omega}\, ,\label{eqdue}
\end{equation}
\begin{equation}
    \mathcal{G}_{P\phi}(\omega)\,\equiv\,\langle\,P\,\phi\,\rangle\,\,=\,\frac{i}{\omega}\, ,\label{e1tre}
\end{equation}
where all the Green functions are computed at zero momentum $k=0$.\\
We are not going to discuss in detail the correlator between the Goldstone operator and the electric current. We just notice that the parameter $\gamma_1$ is finite and will be studied in a separate work. Moreover, the $\mathcal{G}_{P\phi}$ Green function is just the mathematical statement that the operator $\phi$, \textit{i.e.} the phonon, is the Goldstone boson for translations, which are generated by the momentum operator $P$.
We have successfully verified the expressions for the correlators mentioned above. See fig. \ref{figG} for an example. The parameter $\xi$ used in section \ref{sec: spont} is obtained exactly using the $ \langle \phi\phi \rangle$ correlator and matching it to the expression \eqref{eqdue}. More about the physics and the role of $\xi$, especially for the longitudinal sector of the fluctuations, has appeared at the same time of our manuscript in \cite{Ammon:2019apj}.

\bibliographystyle{JHEP}
\bibliography{pseudo}
\end{document}